\documentclass[preprint2]{aastex631}

\usepackage{graphicx}
\usepackage{graphics}
\usepackage{amsmath}
\usepackage{txfonts}
\usepackage{epsfig}
\usepackage{natbib}
\usepackage{times}
\usepackage{amssymb}
\usepackage{color}
\usepackage{longtable}

\newcommand\ergscm{erg\,cm$^{-2}$\,s$^{-1}$}
\newcommand\ergs{erg\,s$^{-1}$}

\newcommand\ctcms{cts\,cm$^{-2}$\,s$^{-1}$}

\newcommand\cmsq{cm$^{-2}$}
\newcommand\integ{{\it{INTEGRAL}}}
\newcommand\fermi{{\it{Fermi}}}
\newcommand\swift{{\it{Swift}}}
\newcommand\xmm{{\it{XMM-Newton}}}
\newcommand\chan{{\it{Chandra}}}
\newcommand\nustar{{\it{NuSTAR}}}
\newcommand\asca{{\it{ASCA}}}
\newcommand\suz{{\it{Suzaku}}}
\newcommand\rxte{{\it{RXTE}}}
\newcommand\hess{{\it{HESS}}}
\newcommand\nh{$N_\mathrm{H}$}
\newcommand\igr{\object{IGR~J16320$-$4751}}

\shorttitle{IGR~J16320$-$4751 with \emph{NuSTAR}}
\shortauthors{Bodaghee et al.}
\graphicspath{{./}{figures/}}

\begin{document}

\title{Drop in the Hard Pulsed Fraction and a Candidate \\ Cyclotron Line in IGR~J16320$-$4751 seen by \emph{NuSTAR}}

\correspondingauthor{A.~Bodaghee}
\email{arash.bodaghee@gcsu.edu}

\author[0000-0002-7315-3732]{A. Bodaghee}
\affil{Dept. of Chemistry, Physics and Astronomy, Georgia College and State University, 231 W. Hancock St., Milledgeville, GA 31061, USA}  

\author{J.-L.~Chiu}
\affil{National Space Organization, National Applied Research Labs, Hsinchu City 300, Taiwan}

\author{J.A. Tomsick},
\affil{Space Sciences Laboratory, University of California, 7 Gauss Way, Berkeley, CA 94720, USA}

\author{V. Bhalerao} 
\affil{Indian Institute of Technology Bombay, Powai, Mumbai 400076, India}  

\author{E. Bottacini}
\affil{W.W. Hansen Experimental Physics Laboratory \& Kavli Institute for Particle Astrophysics and Cosmology, Stanford University, 452 Lomita Mall, Palo Alto, CA 94305, USA}
\affil{Dipartimento di Fisica e Astronomia ``G. Galilei,'' Universit\`{a} di Padova, Via Belzoni 7, 35131 Padova, Italy}
\affil{Eureka Scientific, 2452 Delmer Street, Suite 100, Oakland, CA 94602, USA}

\author{M. Clavel},
\affil{Universit\'{e} Grenoble Alpes, CNRS, IPAG, 38000 Grenoble, France}

\author{C. Cox}
\affil{Dept. of Chemistry, Physics and Astronomy, Georgia College and State University, 231 W. Hancock St., Milledgeville, GA 31061, USA}  

\author{F. F\"{u}rst}
\affil{Quasar Science Resources SL for ESA, European Space Astronomy Centre (ESAC), Science Operations Department, 28692 Villanueva de la Ca\~{n}ada, Madrid, Spain}

\author{M.J. Middleton}  
\affil{Department of Physics and Astronomy, University of Southampton, Highfield, Southampton SO17 1BJ, UK}

\author{F. Rahoui}  
\affil{European Southern Observatory, Karl Schwarzchild-Strasse 2, 85748 Garching bei Munchen, Germany}

\author{J. Rodriguez}
\affil{Universit\'{e} Paris-Saclay, Universit\'{e} Paris Cit\'{e}, CEA, CNRS, AIM, 91191, Gif-sur-Yvette, France}

\author{P. Romano}
\affil{INAF, Osservatorio Astronomico di Brera, Via E. Bianchi 46, 23807 Merate, Italy}  

\author{J. Wilms} 
\affil{Dr. Karl-Remeis-Sternwarte and ECAP, Sternwartstrasse 7, 96049 Bamberg, Germany}





\begin{abstract}

We report on a timing and spectral analysis of a 50-ks \nustar\ observation of \igr\ ($=$ \object{AX~J1631.9$-$4752}); a high-mass X-ray binary hosting a slowly-rotating neutron star. In this observation from 2015, the spin period was 1,308.8$\pm$0.4 s giving a period derivative $\dot{P} \sim 2 \times 10^{-8}$ s s$^{-1}$ when compared with the period measured in 2004. In addition, the pulsed fraction decreased as a function of energy, as opposed to the constant trend that was seen previously. This suggests a change in the accretion geometry of the system during the intervening 11 years. The phase-averaged spectra were fit with the typical model for accreting pulsars: a power law with an exponential cutoff. This left positive residuals at 6.4 keV attributable to the known iron K$\alpha$ line, as well as negative residuals around 14 keV from a candidate cyclotron line detected at a significance of 5$\sigma$. We found no significant differences in the spectral parameters across the spin period, other than the expected changes in flux and component normalizations. A flare lasting around 5 ks was captured during the first half of the observation where the X-ray emission hardened and the local column density decreased. Finally, the binary orbital period was refined to 8.9912$\pm$0.0078 d thanks to \swift/BAT monitoring data from 2005--2022.

\end{abstract}



\section{Introduction} 
\label{sec:intro}

Hard X-ray monitoring of the Galactic Plane by \integ\ has uncovered dozens of new High-Mass X-ray Binaries \citep[HMXBs:][]{wal15,sid18,kre19}. These are systems in which a neutron star (NS) or a black hole (BH) accretes from a massive companion star ($M \gtrsim 5 \,\, M_{\odot}$). Given that \integ's position uncertainty is a few arcminutes, the only way to tell that these objects are HMXBs is thanks to follow-up observations with X-ray telescopes such as \chan, \nustar, \suz, \swift, and \xmm. 

During follow-up observations, many of these HMXBs presented characteristics that understandably hindered their detection in lower-energy X-ray surveys. They could be extremely obscured below 5 keV with X-ray absorbing column densities (\nh) several times $10^{23}$ \cmsq, or an order of magnitude more than the cumulative Galactic absorption along the line of sight \citep[e.g.,][]{mat03,wal03,pat04}. Some stayed at a low-level of emission for months or years ($\sim 10^{32}$ \ergs) and awakened with huge flares where the luminosity would increase by up to 6 orders of magnitude \citep[i.e., supergiant fast X-ray transients or SFXTs:][]{int05,neg06,rom14,boz15,rom15,sid16}. Finally, they could have spin periods lasting around 1 ks \citep[e.g.,][]{zur06,bod06,rod06}.

\begin{figure*}[]
\begin{center}
\includegraphics[width=\textwidth]{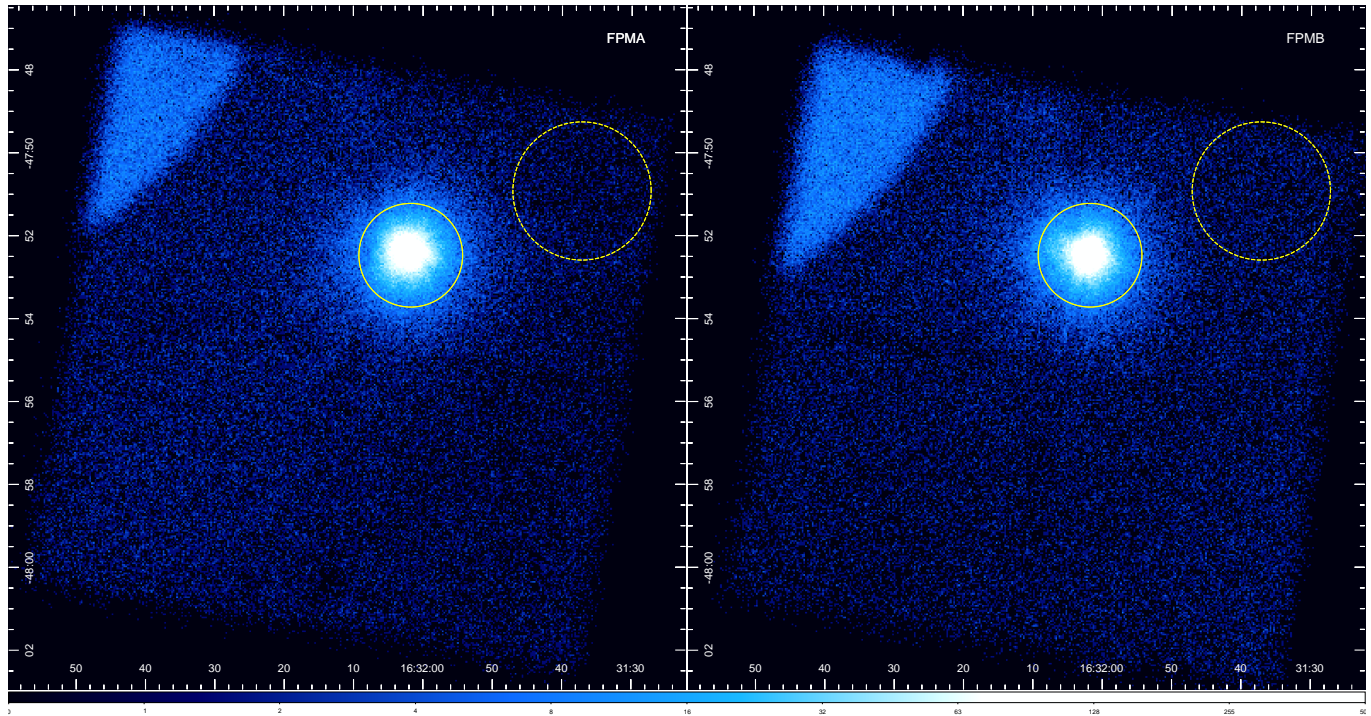}
\end{center}
\caption{Images of IGR~J16320$-$4751 gathered with \nustar\ FPMA (\emph{left}) and FPMB (\emph{right}) in 3--79\,keV. The images are presented in J2000.0 equatorial coordinates, they are scaled logarithmically, and extraction regions for the source (75$^{\prime\prime}$-radius) and background (100$^{\prime\prime}$-radius) are indicated.}
\label{fig:image}
\end{figure*}

\begin{deluxetable*}{ l c c c c c c }
\tablewidth{0pt}
\tabletypesize{\scriptsize}
\tablecaption{Journal of observations of IGR~J16320$-$4751. \label{tab:log}}
\tablehead{
\colhead{telescope} & \colhead{observation ID} & \colhead{pointing R.A. (J2000)} & \colhead{pointing decl. (J2000)}	& \colhead{start date (UT)} 	& \colhead{end date (UT)} 	&  \colhead{effective exposure (ks)} }
\startdata
\nustar\ 		& 30001008002	& 248.0277 & $-$47.904 		& 2015-06-06 18:46:07 	& 2015-06-07 23:36:07 & 49.845	\\
\swift/XRT		& 00081642001	& 247.9418 & $-$47.8672		& 2015-06-06 20:07:38 	& 2015-06-06 22:06:56 & 1.703 
\enddata
\end{deluxetable*}

One of the obscured HMXBs detected early in the \integ\ mission was \igr\ \citep{tom03,rod03}. This turned out to be the high-energy counterpart of an unclassified X-ray source named AX J1631.9$-$4752 that was discovered two years earlier by \asca\ \citep{sug01}. A coherent pulsation of 1,309$\pm$40 s, consistent with the spin period of an accreting NS, was measured with \xmm\ \citep{lut05}, and was later refined to 1,303.8$\pm$0.9 s \citep{rod06}. While the source varies on short timescales, its average, hard X-ray flux since its discovery by \asca\ has stayed within a narrow range \citep{kri13,kri22}. Near-IR spectroscopy suggests a supergiant donor star whose spectral type is BN0.5 Ia \citep{col13}, thereby confirming its status as an HMXB at a distance of $\sim$3.5 kpc \citep{rah08}. Six arcminutes away from \igr\ is an unrelated pulsar wind nebula \citep{ace15} detected in the gamma-rays by \fermi/LAT \citep[\object{4FGL~J1633.0$-$4746e}:][]{abd20} and \hess\ \citep[\object{HESS~J1632$-$478}:][]{aha06}.

With a short orbital period of 8.99$\pm$0.01 days \citep{cor05,lev11,gar18}, the large absorption (\nh $\gtrsim10^{23}$ \cmsq) is likely due to the NS being shrouded in the wind of its companion star. In the case of supergiant donors, the stellar wind is the main contributor to the photoelectric absorption and the Fe K$\alpha$ emission \citep{gim15,pra18}. In the specific case of \igr, \citet{gar18} demonstrated that these parameters as measured by \xmm\ were modulated with the orbital period of the binary as determined by \swift/BAT, i.e., they were related to the configuration of the X-ray source and surrounding wind as viewed by the observer. 

In 2015, \nustar\ pointed at \igr\ for 29 hours and those results are presented for the first time in this paper. Procedures for analysis of the \nustar\ data are described in Section \ref{sec:ObsDAn}. Results from timing and spectral analyses are in Section \ref{sec:results}, and they are discussed in Section \ref{sec:discussion}.

\section{OBSERVATIONS AND DATA ANALYSIS} \label{sec:ObsDAn}

\nustar\ observed \igr\ for a total of 103.6 ks from 18:46:07 (UT) on 2015 June 6, until 23:36:07 (UT) on 2015 June 7. This observation (ObsID: 30101026002) collected 49.8 ks of on-source time (Good Time Intervals or GTIs) on each of its two focal plane modules A and B (FPMA and FPMB) which have a 13'$\times$13' field of view (FoV). 

The \nustar\ data were reduced using {\tt nupipeline} and {\tt nuproducts} from the \nustar\ Data Analysis Software ({\tt NuSTARDAS} v2.1.1) as distributed with {\tt HEASoft} \citep[v6.29:][]{hea14}. Response files were linked to the latest calibration database available at the time ({\tt CALDB} 20220105). We extracted source counts for FPMA and FPMB from a circular region with a radius of 75$^{\prime\prime}$ centered at the source position listed in the 4XMM Serendipitous Source Catalog \citep{web20}: R.A. (J2000.0) $= 16^{\mathrm{h}}32^{\mathrm{m}}01.76^{\mathrm{s}}$, and Decl. = $-47^{\circ}52^{\prime}29.0^{\prime\prime}$. Background counts were extracted from a circular region (100$^{\prime\prime}$ radius) situated away from the source region while remaining on the same detector chip. 

Figure\,\ref{fig:image} presents the field of \igr\ from FPMA and FPMB along with the source and background extraction regions. Stray light from \object{GX~340$+$0}, situated 3$^{\circ}$ outside the FoV, affected a small portion of one detector chip from each module. The boundary of the stray light contamination region was far enough away from the location of \igr\ that it did not impact our analysis. 

A concurrent \swift/XRT observation (ObsID: 00081642001) was performed on 2015 June 6 between 20:07:38 and 22:06:56 (UT) for an effective exposure time of 1,703 s. The \swift/XRT data were processed according to the standard procedure of {\tt xrtpipeline} (v0.13.6). Source counts were extracted from inside a circular region whose radius was 20 pixels (1 pixel $\sim$2.36$^{\prime\prime}$). Background events were taken from an annular source-free region centered on the source (inner/outer radii of 70/110 pixels). Ancillary response files were generated with {\tt xrtmkarf} (v0.6.3) to account for the different extraction regions, vignetting, and PSF corrections, while the spectral redistribution matrix was the most recent version available (20130101v014). 

Light curve data for \igr\ was downloaded from the \swift/BAT Hard X-ray Transient Monitor$\footnote{https://swift.gsfc.nasa.gov/results/transients/}$ \citep{kri13} where the source is listed under its alias AX~J1631.9$-$4752. These data consist of a count rate and error in 15--50 keV collected during each orbital pointing of the \swift\ satellite between 2005 February 14 and 2022 July 20. Data of poor quality were excluded by selecting only those rows where ``\texttt{DATA\_FLAG==0}.'' The remaining 78,882 pointings have exposure times that range from 64 s to 2.64 ks (an average of 665 s) giving a total effective exposure time of 52.5 Ms spread over 550 Ms of calendar time (6,366 d).

\begin{figure}[]
\begin{center}
\includegraphics[width=0.45\textwidth]{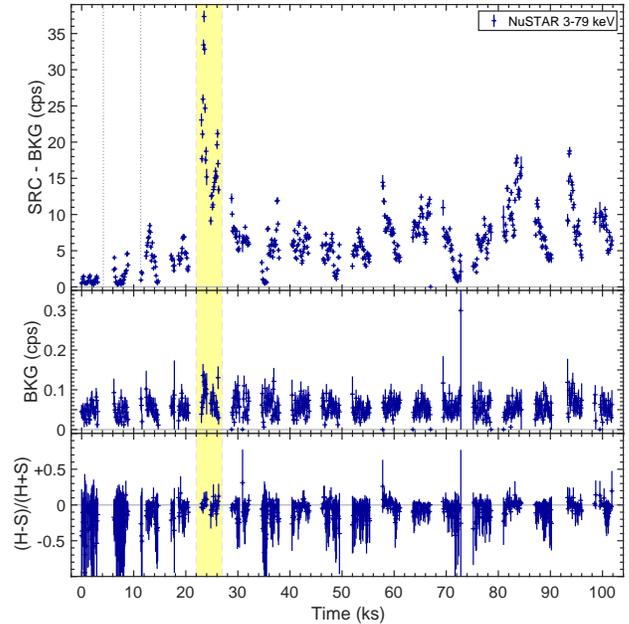}
\end{center}
\caption{Light curve and hardness ratio from \nustar\ (3--79 keV) for \igr\ with 100 s in each bin ($T_{0} =$ MJD 57179.795322). The total (FPMA$+$FPMB) net source count rate is presented in the top panel where the background count rate, as shown in the middle panel, has been scaled by area and subtracted. The hardness ratio is featured in the bottom panel where $S$ and $H$ represent count rates in 3--10 keV and 10--79 keV, respectively. The flare corresponds to an epoch 22--27 ks into the observation (highlighted in light yellow and bounded by dashed lines). The dotted lines indicate the start and stop times of the contemporaneous \swift/XRT pointing.}
\label{fig:light}
\end{figure}

Relying on the 4XMM position above, we performed barycentric corrections on the \nustar\ data in {\tt nuproducts}, while the \swift\ XRT and BAT data were barycentered using \texttt{barycorr} with the orbital ephemeris parameter set to \texttt{geocenter}. Timing and spectral data were analyzed in {\tt Xronos} (v6.0) and {\tt XSpec} \citep[v12.12.0:][]{arn96}, respectively, where the latter assumed elemental abundances from \citet{wil00} and photon-ionization cross-sections from \citet{ver96}. \nustar\ data were restricted to 3--79 keV, \swift/XRT data were limited to 0.3--10 keV, and known bad channels from both telescopes were ignored. \nustar\ spectral counts were grouped such that each bin had a minimum significance of 5$\sigma$ (for phase-resolved analysis) and 10$\sigma$ (for phase-averaged analysis) permitting the use of $\chi^{2}$ statistics, while \swift/XRT spectral data were grouped at 20 counts per bin. Unless specified otherwise, error bars in the figures indicate 1$\sigma$ confidence boundaries, while error values cited in the text and tables are given at 90\% confidence. A journal of \nustar\ and \swift/XRT observations is provided in Table\,\ref{tab:log}.

\begin{figure}[]
\begin{center}
\includegraphics[width=0.45\textwidth]{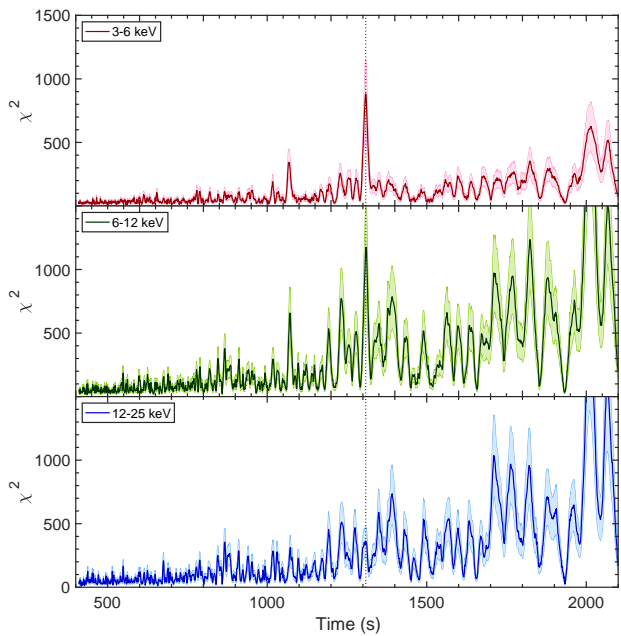}
\end{center}
\caption{Spin period search ($\chi^{2}$ distribution) on the \nustar\ light curve of IGR J16320$-$4751 with 20 bins per period and a resolution of 1.2 s in three energy bands: 3--6 keV (top panel; red curve); 6--12 keV (middle panel; green curve); and 12--25 keV (bottom panel; blue curve). The best-fitting spin period ($P= $ 1,308.8 s) is denoted by a dotted line.}
\label{fig:spin_search}
\end{figure}

\section{RESULTS} \label{sec:results}

\subsection{Timing Analysis} \label{subsec:TimingAnalysis}

\subsubsection{Light Curve} \label{subsubsec:LC}

Figure \ref{fig:light} shows the \nustar\ background-subtracted light curve (3--79 keV) and hardness ratio where net source counts from both modules have been summed. The hardness ratio is defined as $(H-S)/(H+S)$ where $S$ (3--10 keV) and $H$ (10--79 keV) are net count rates. In FPMA, there were a total of 110,296$\pm$333 net counts in 49.89 ks of effective exposure time, and in FPMB, there were 98,064$\pm$314 net counts in 49.84 ks. Summing the net counts from both modules returned a total of 208,360$\pm$458 counts or a count rate of 4.18$\pm$0.01 counts s$^{-1}$.

\begin{figure}[]
\begin{center}
\includegraphics[width=0.45\textwidth]{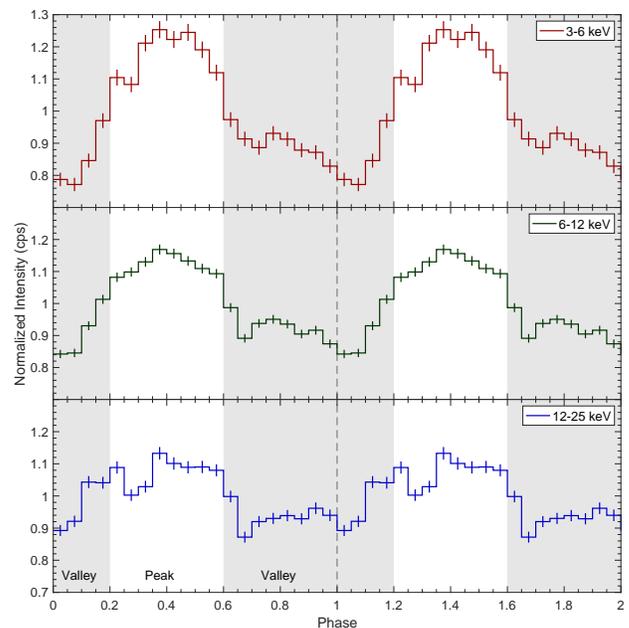}
\end{center}
\caption{Pulse profiles ($P =$ 1,308.8 s) in three energy bands from the \nustar\ observation. The pulse is repeated once for ease of viewing. Bins attributed to the peak (white; phases 0.2--0.6) and to the valley (gray; phases 0.0--0.2 and 0.6--1.0) are designated for phase-resolved spectroscopy. All panels have the same vertical scale.}
\label{fig:spin_profile}
\end{figure}

A flare lasting $\sim$5 ks was noticed around 22 ks into the observation. This time interval of 22--27 ks after the observation began was called the ``flare'' epoch. During the flare, there were 14,321$\pm$120 net counts in FPMA, and 12,756$\pm$113 net counts in FPMB, with an effective exposure time of 2.26 ks per module. When the modules were summed, the total net counts was 27,078$\pm$165 with an average count rate of 11.98$\pm$0.07 counts s$^{-1}$. 

The rest of the observation, i.e., excluding the flare, was referred to as the ``non-flare'' epoch. This epoch contained 95,975$\pm$311 net counts in FPMA (47.64 ks), and 85,308$\pm$293 net counts in FPMB (47.59 ks). A sum of both modules gave 181,283$\pm$427 total net counts or 3.81$\pm$0.01 counts s$^{-1}$.

In \swift/XRT, there were 112$\pm$12 net counts in 1,703 s for a count rate of (6.6$\pm$0.7)$\times10^{-2}$ counts s$^{-1}$.

\begin{figure}[]
\begin{center}
\includegraphics[width=0.45\textwidth]{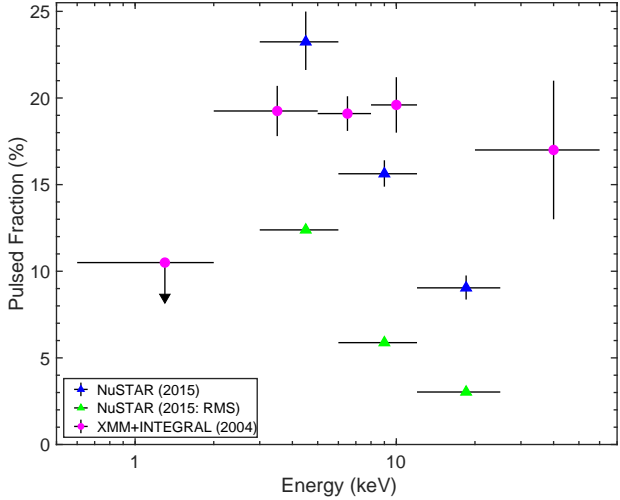}
\end{center}
\caption{Energy dependence of the pulsed fraction. The pulsed fractions of IGR J16320$-$4751 light curves are shown for this 2015 \nustar\ observation (blue; $P=1,308.8$ s) and for the 2004 observation by \citet{rod06} combining \xmm\ and \integ\ (magenta; $P=1,303$ s). Root mean square (RMS) values of the \nustar\ light curves (green) are also plotted as a reference.}  
\label{fig:PF_E}
\end{figure}

\subsubsection{Spin Period} \label{subsubsec:SpinP}

A periodicity search was performed with \texttt{efsearch} on the source ($+$ background) light curve (0.1-s resolution) of the full observation in five energy bands: 3--6 keV, 6--12 keV, 12--25 keV, 25--79 keV, and 3--79 keV. A coherent pulsation near the known period of 1,300 s was detected in all energy bands except 25--79 keV. The pulsation was detected most significantly at lower energies (Fig.\,\ref{fig:spin_search}). Figure\,\ref{fig:spin_profile} presents the pulse profile with 20 bins per period for energies up to 25 keV. In the 3--6-keV band, the best-fitting period was 1,308.8$\pm$0.4 s where the centroid was determined with the \citet{pre89} fast algorithm for Lomb-Scargle periodograms \citep{lom76,sca82} and the error from \citet{hor86} and \citet{lea87}. 

Featuring a single broad peak and a mirrored valley, the shape of the pulse profile from this 2015 \nustar\ observation is similar to the one from 2004 using \xmm\ and \integ\ \citep{rod06}. However, there was a significant difference in the pulsed fraction between the observations. The pulsed fraction is defined as $(I_{\mathrm{max}}-I_{\mathrm{min}})/(I_{\mathrm{max}}+I_{\mathrm{min}})$ where $I_{\mathrm{max}}$ and $I_{\mathrm{min}}$ are the normalized count rates in the highest (phase: 0.35--0.40) and lowest intensity bins (phase: 0.0--0.05), respectively. Figure\,\ref{fig:PF_E} shows that the pulsed fraction decreased significantly as a function of energy during this \nustar\ observation. The root mean square (RMS) of the \nustar\ light curves had a similar negative correlation with energy. This behavior is different from what was previously seen by \xmm\ and \integ\ where the pulsed fraction was consistent with being constant as a function of energy \citep{rod06}. 

\subsubsection{Orbital Period} \label{subsubsec:OrbitalP}

\begin{figure}[]
\begin{center}
\includegraphics[width=0.45\textwidth]{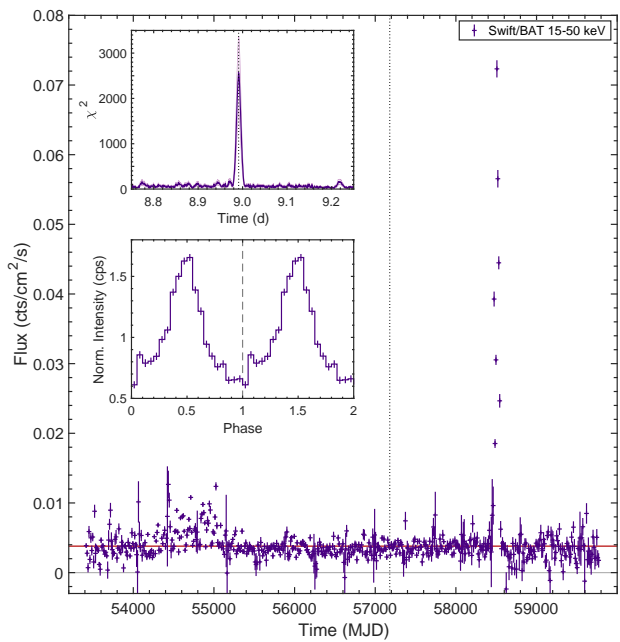}
\end{center}
\caption{Light curve and orbital period search from \swift/BAT Transient Monitor \citep{kri13} data of IGR J16320$-$4751 in 15--50 keV. The main panel presents count rates where each bin collects 1 Ms of exposure time. The average count rate ($3.8\times10^{-3}$ \ctcms), which excludes the prominent flare (whose apex occurs on MJD 58505), is denoted as a horizontal red line. The dotted line indicates the time of the \nustar\ observation. The upper inset panel shows results from an orbital period search centered at 8.9912 day (dotted vertical line), with 20 bins per period and a resolution of 50 s, while the lower inset panel gives two cycles of the orbital period. }    
\label{fig:lc_bat}
\end{figure}

The 17-year BAT light curve of \igr\ illustrates the stability of the source flux on timescales of years (main panel of Fig.\,\ref{fig:lc_bat}). The figure reveals a prominent flare around MJD 58500 where, over the course of 2--3 months, the source count rate increased up to a factor of nearly 20 to $7.4 \times 10^{-2}$ \ctcms\ from a mean value (without the flare) of $3.8 \times 10^{-3}$ \ctcms. We used \texttt{efsearch} with 20 bins per period and a resolution of 50 s to generate a periodogram (upper inset panel of Fig.\,\ref{fig:lc_bat}), and we fit its peak with a Gaussian to obtain an orbital period of 8.9912$\pm$0.0078 d with $T_{\phi_{0}}=$ MJD 59760.449555 corresponding to the phase bin with the lowest flux in the folded light curve (lower inset panel of Fig.\,\ref{fig:lc_bat}). This means the \nustar\ observation coincided with orbital phases 0.97--1.0 and 0.0--0.11. The orbital period matches the value obtained by \citet{gar18} at a higher significance thanks to 5 additional years of data.

\subsection{Spectral Analysis} \label{subsec:SpecAnalysis}

\subsubsection{Phase-averaged Spectroscopy}

The spectral data from FPMA and FPMB were collected so that each bin had a signal-to-noise ratio ($S/N$) of at least 10. These spectra were initially fit with a power law attenuated by a photoelectric absorption component at low energies (\texttt{Tbabs}) and an exponential cutoff at high energies (\texttt{CutoffPL}). An instrumental constant (\texttt{Const}) was fixed at 1 for FPMA and allowed to vary for FPMB. This constant was 0.95$\pm$0.01 in all cases except for the flare epoch where it was 0.96$\pm$0.02. This spectral model is called ``Model 1'' (M1), and it leaves positive residuals around 6.4 keV attributable to an iron K$\alpha$ line often seen in HMXBs with supergiant donors. 

A Gaussian to account for the iron line was then included as ``Model 2'' (M2). The improvement in fit quality was significant with $\chi_{\nu}^{2}$/d.o.f. dropping from 1.40/930 in M1 to 1.16/927 in M2 for the full observation. A similar improvement was seen in the non-flare epoch with $\chi_{\nu}^{2}$/d.o.f. decreasing from 1.35/873 in M1 to 1.13/870 in M2. 

The addition of a cyclotron line \citep[\texttt{cyclabs}:][]{mih90,mak90} in ``Model 3'' (M3) to account for negative residuals in 10--20 keV led to a slight decrease in $\chi_{\nu}^{2}$/d.o.f. to 1.12/924 for the full observation. For the non-flare spectrum, the inclusion of the cyclotron component led to a significant improvement in the fit quality with $\chi_{\nu}^{2}$/d.o.f. reduced to 1.07/867.

\begin{figure*}
\begin{center}
\includegraphics[width=0.45\textwidth]{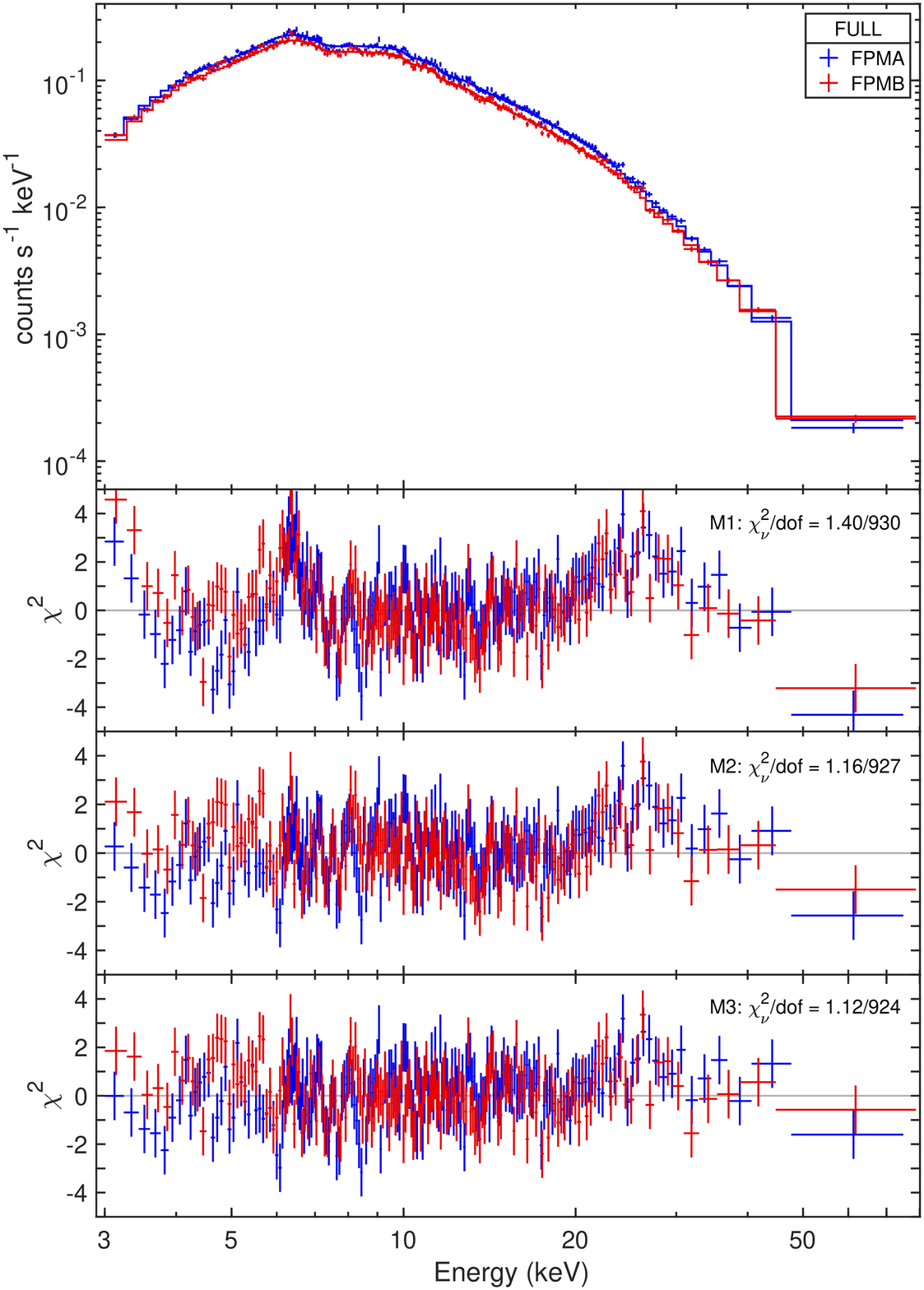}\hspace{5mm}\includegraphics[width=0.45\textwidth]{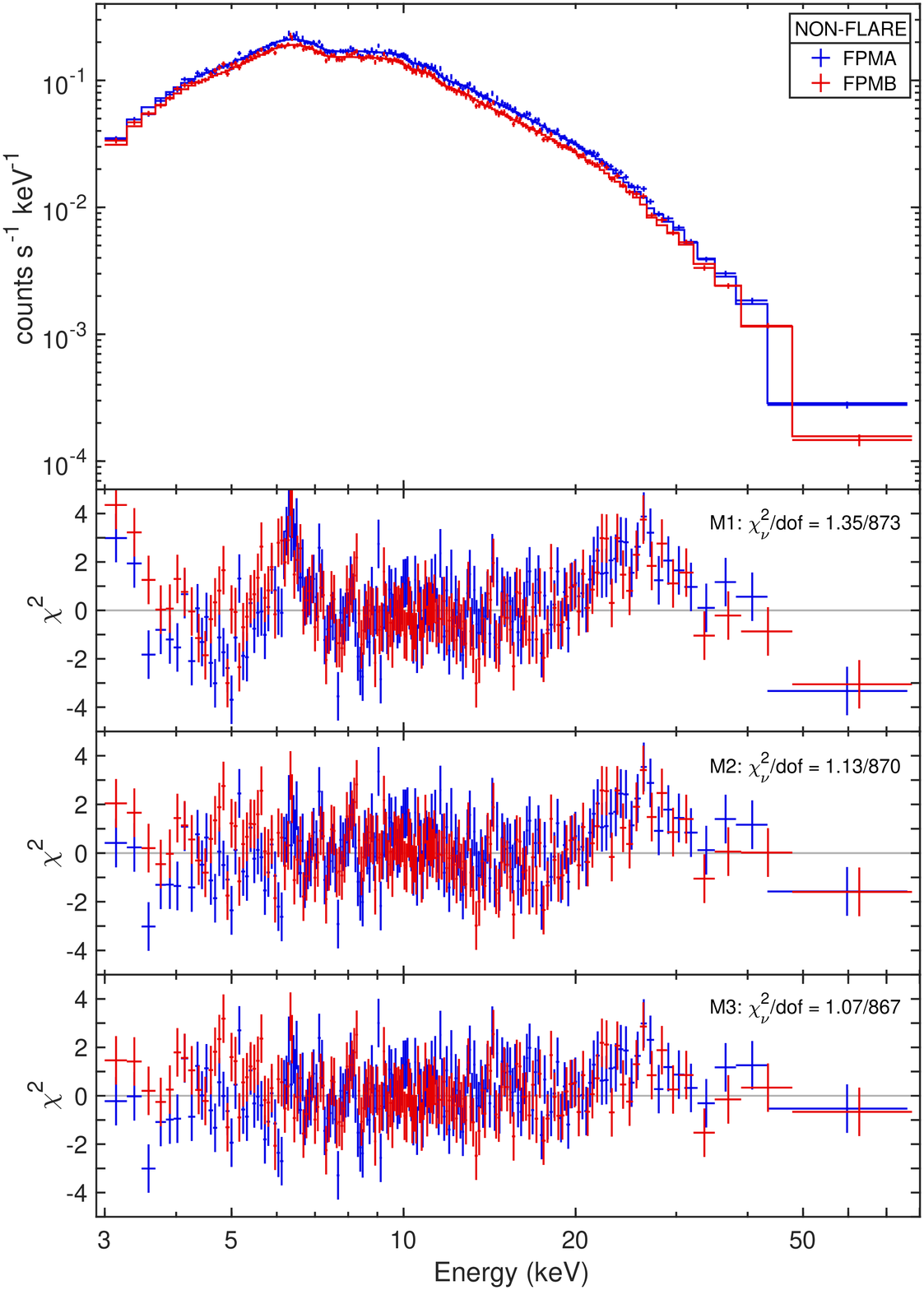}
\end{center}
\caption{Phase-averaged \nustar\ spectra for IGR J16320$-$4751. The column of panels on the left presents the background-subtracted source spectrum from the full observation, while the column of panels on the right features the spectrum restricted to the non-flare epoch as defined in Fig.\,\ref{fig:light}. Within each column, the top panel contains the spectral data and the best-fitting model (M3), while the lower rows of panels show residuals from fitting the three {\tt CutoffPL}-based models listed in Table \ref{table_fit}. Each bin collects a minimum significance of 20$\sigma$ to better highlight deviations from the model (compared with 10$\sigma$ during fits).}
\label{fig:spec_multi}
\end{figure*}

\begin{deluxetable*}{l | ccc | cc | cc |}
\tablecaption{Parameters from different empirical models fit to \nustar\ spectra of IGR J16320$-$4751. }
\tablehead{
\colhead{} & \colhead{Flare} & \colhead{Non-Flare}  & \colhead{Full} \vline & \colhead{Non-Flare} &  \colhead{Full} \vline & \colhead{Non-Flare} &  \colhead{Full} \\ 
\cline{2-8}
\colhead{} & \multicolumn{3}{c|}{Model 1}  & \multicolumn{2}{c|}{Model 2} & \multicolumn{2}{c}{Model 3} 
} 
\startdata
     &     \multicolumn{7}{c|}{photoelectric absorption ({\tt Tbabs}) }     
\\   
$N_{\mathrm{H}}^{a}$ 
& 11.7$_{-2.5}^{+2.6}$ 	
& 19.4$\pm$0.9
& 18.0$\pm$0.9 
& 13.4$\pm$1.2 
& 12.2$\pm$1.1 
& 10.5$\pm$1.6
& 10.4$\pm$1.4 
   \\  
      &     \multicolumn{7}{c|}{cutoff power law ({\tt CutoffPL})}     
\\
$\Gamma$ 
& $-$0.04$\pm$0.12
& 0.48$\pm$0.04 
& 0.39$\pm$0.04 
& 0.18$\pm$0.06 
& 0.09$\pm$0.06
& $-0.08_{-0.11}^{+0.10}$ 
& $-0.09_{-0.10}^{+0.09}$  
\\
$E_{\mathrm{cut}}^{b}$   
& 11.6$_{-0.8}^{+1.0}$ 
& 15.2$_{-0.5}^{+0.6}$ 
& 14.3$_{-0.4}^{+0.5}$ 
& 12.7$\pm$0.5 
& 12.1$\pm$0.4 
& 11.0$\pm$0.6 
& 10.9$_{-0.5}^{+0.6}$ 
\\
$N_{\mathrm{cut}}^{c}$   
& 37.9$_{-7.6}^{+9.5}$ 
& 34.4$_{-2.7}^{+2.9}$ 
& 31.7$_{-2.3}^{+2.5}$ 
& 18.5$_{-2.2}^{+2.4}$ 
& 17.2$_{-1.9}^{+2.0}$ 
& 11.9$_{-2.1}^{+2.4}$ 
& 12.6$_{-2.1}^{+2.4}$ 
\\
      &     \multicolumn{7}{c|}{Fe K$_{\alpha}$ line ({\tt Gauss})}     
\\
$E_{\mathrm{Fe}}^{b}$   
& \nodata  
& \nodata  
& \nodata
& 6.34$\pm$0.06 
& 6.34$\pm$0.06  
& 6.28$_{-0.07}^{+0.06}$  
& 6.30$_{-0.07}^{+0.08}$  
\\
$\sigma_{\mathrm{Fe}}^{b}$   
& \nodata  
& \nodata  
& \nodata  
& 0.44$\pm$0.08 
& 0.46$\pm$0.08  
& 0.50$\pm$0.08 
& 0.48$\pm$0.09 
\\
$N_{\mathrm{Fe}}^{c}$   
& \nodata  
& \nodata  
& \nodata  
& 1.96$_{-0.32}^{+0.34}$ 
& 2.12$_{-0.33}^{+0.36}$ 
& 2.31$_{-0.40}^{+0.43}$ 
& 2.27$_{-0.41}^{+0.43}$  
\\
      &     \multicolumn{7}{c|}{CRSF (\texttt{cyclabs})}     
\\
$E_{\mathrm{cyc}}^{b}$   
& \nodata  
& \nodata  
& \nodata 
& \nodata
& \nodata  
& 14.3$_{-1.2}^{+0.9}$  
& 13.4$_{-2.0}^{+1.0}$  
\\
$\sigma_{\mathrm{cyc}}^{b}$   
& \nodata  
& \nodata  
& \nodata  
& \nodata 
& \nodata  
& 4.2$_{-1.2}^{+1.4}$  
& 4.3$_{-1.4}^{+2.1}$
\\
$\tau_{\mathrm{cyc}}^{d}$   
& \nodata  
& \nodata  
& \nodata  
& \nodata 
& \nodata  
& 0.10$\pm$0.02
& 0.08$_{-0.02}^{+0.03}$
\\  
\cline{1-8}
$\chi_{\nu}^{2}$ (d.o.f.)  
& 1.17 (224) 
& 1.35 (873) 
& 1.40 (930) 
& 1.13 (870) 
& 1.16 (927) 
& 1.07 (867)   
& 1.12 (924) 
\\   
$F^{e}$ 
& 26.46$_{-1.15}^{+0.33}$
& 8.85$_{-0.08}^{+0.05}$ 
& 9.65$_{-0.07}^{+0.06}$
& 8.91$_{-0.11}^{+0.05}$ 
& 9.72$_{-0.10}^{+0.05}$
& 8.97$_{-0.30}^{+0.04}$   
& 9.78$_{-0.27}^{+0.04}$   
\\
\enddata
\tablecomments{ (a) equivalent hydrogen column density ($\times 10^{22}$ cm$^{-2}$); (b) in keV; (c) normalization at 1 keV ($\times10^{-4}$ ph s$^{-1}$ cm$^{-2}$ keV$^{-1}$); (d) optical depth; (e) model-derived flux in the 2--10 keV band ($\times10^{-11}$ ergs s$^{-1}$ cm$^{-2}$) }   
\label{table_fit}
\end{deluxetable*}

\begin{figure}[]
\begin{center}
\includegraphics[width=0.45\textwidth]{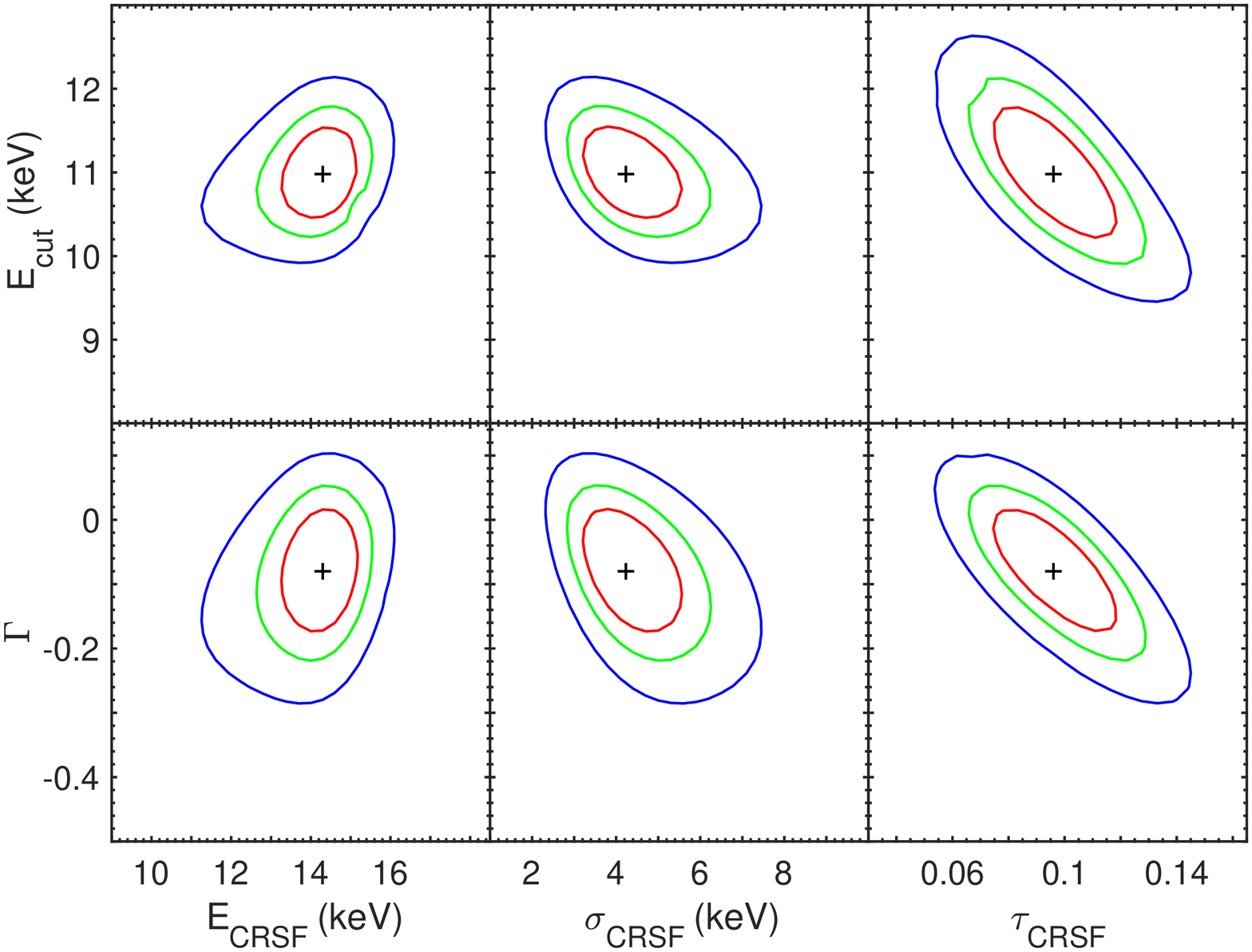}
\end{center}
 \caption{Confidence regions for the cyclotron parameters from the best-fitting spectral model (M3) during the non-flare epoch. Red, green, and blue lines represent 68\%, 90\%, and 99\% confidence contours, respectively, around the optimal value shown as a cross.}   
\label{fig:steppar}
\end{figure}

Figure \ref{fig:spec_multi} presents the \nustar\ spectra from the full observation (left column) and from the non-flare epoch (right column) with the best-fitting model (M3), as well as the residuals from models that gradually include more components. The model parameters are listed in Table \ref{table_fit}. 

During the non-flare epoch, the best-fitting spectral parameters from M3 were a column density \nh\ $=$ (10.5$\pm$1.6)$\times10^{22}$ \cmsq, a photon index $\Gamma = -0.08_{-0.11}^{+0.10}$ with a cutoff energy $E_{\mathrm{cut}} = 11.9_{-2.1}^{+2.4}$ keV. The energy and width of the iron line were $6.28_{-0.07}^{+0.06}$ keV and 0.50$\pm$0.08 keV, respectively. The cyclotron line had a centroid energy of $E_{\mathrm{cyc}} = 14.3_{-1.2}^{+0.9}$ keV with a width of $\sigma = 4.2_{-1.2}^{+1.4}$ keV and a low optical depth $\tau =$ 0.10$\pm$0.02. Confidence contours for the continuum and CRSF parameters are shown in Fig.\,\ref{fig:steppar}.

Models with a high-energy cutoff \citep[\texttt{highecut}:][]{whi83}, a negative and positive power-law exponential \citep[\texttt{NPEX}:][]{mak99}, a reflection component \citep[\texttt{reflect}:][]{mag95}, thermal Comptonization \citep[\texttt{compTT} and \texttt{nthComp}:][]{tit94,zdz96,zyc99}, and a Fermi-Dirac cutoff \citep[\texttt{FDcut}:][]{tan86} were also attempted. None of them provided a significant improvement over M3. 

There were too few counts in \swift/XRT ($\sim$100) to be useful for our spectral analysis. Fitting the \swift\ 0.5--10-keV spectrum by itself with an absorbed power law led to unconstrained parameters, whether grouping to minimum of 20 counts per bin, or when leaving the counts unbinned and using \citet{cas79} statistics. Jointly fitting the \swift\ and \nustar\ spectra returned an instrumental constant $\sim$4 for \nustar, even during the non-flare epoch. Depending on the epoch, fits to the combined \swift\ and \nustar\ spectra in 0.5--79 keV gave a column density of (16--21)$\times 10^{22}$ \cmsq. Since this was consistent with the values we obtained when relying on \nustar\ alone, as well as with the values cited by \citet{rod06} and by \cite{gar18}, the \swift\ spectral data were no longer included in the analysis.

\begin{figure*}[]
\begin{center}
\includegraphics[width=0.45\textwidth]{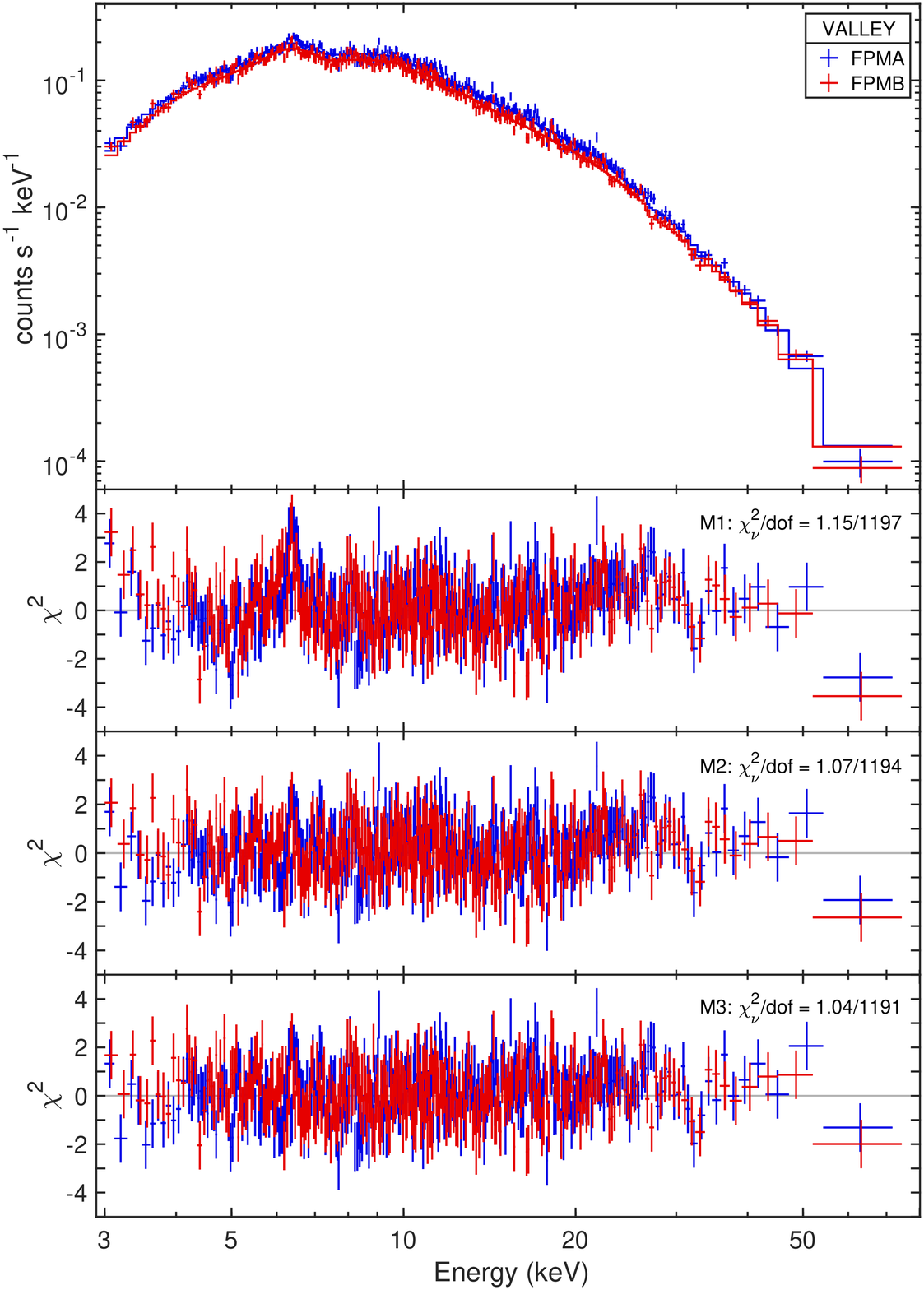}\hspace{5mm}\includegraphics[width=0.45\textwidth]{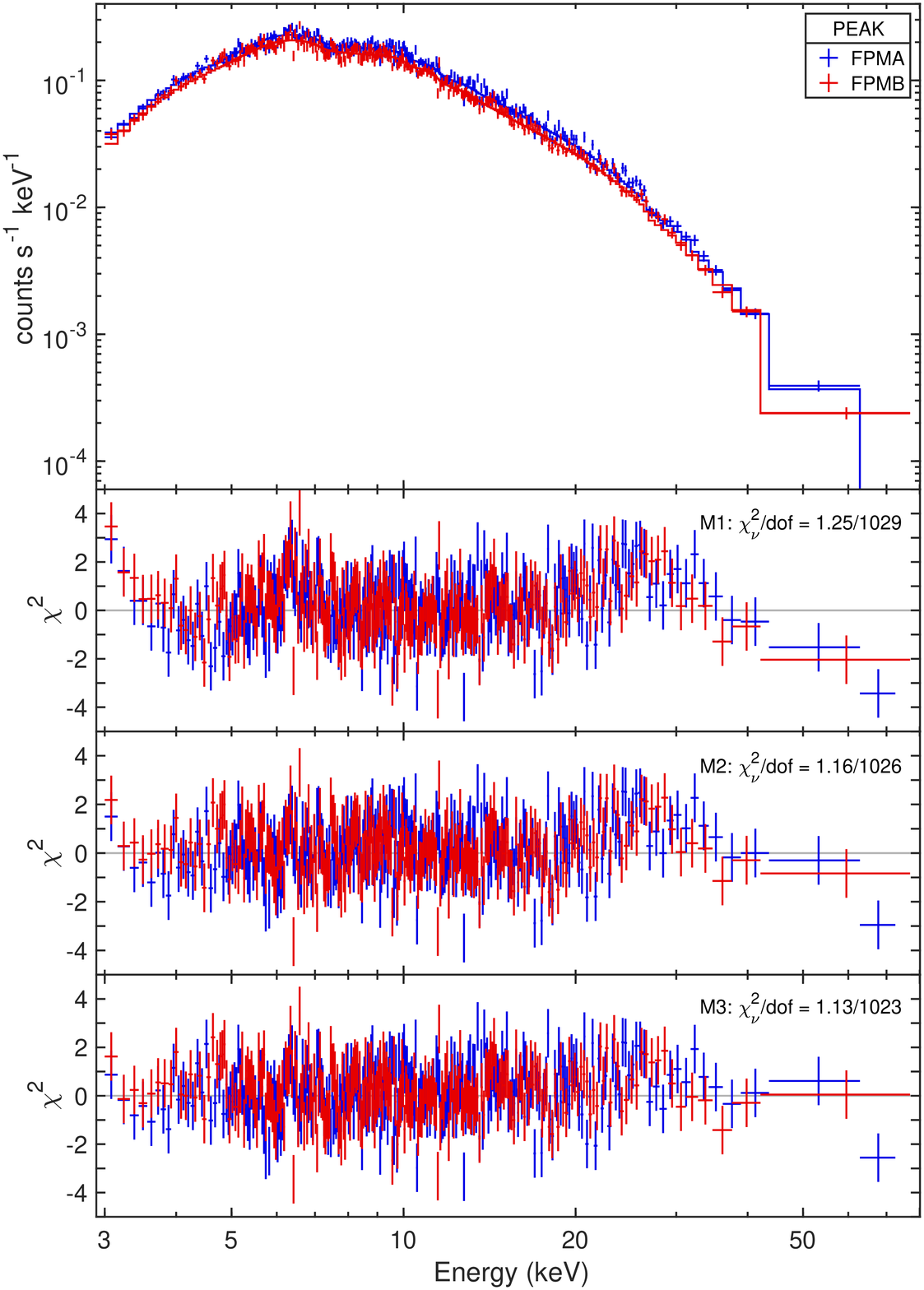}
\end{center}
 \caption{Phase-resolved \nustar\ spectra for IGR J16320$-$4751 from FPMA (blue) and FPMB (red) during the non-flare epoch. The column of panels on the left shows the source spectrum during the Valley phases defined in Fig.\,\ref{fig:spin_profile}, while the column of panels on the right shows data from the Peak phase. In each column, the top row shows the spectrum and best-fitting model (M3), while the second, third, and fourth rows present residuals from fitting the models listed in Fig.\,\ref{table_PR_fit}. For visual clarity, the spectra were rebinned to a minimum significance of 10$\sigma$ (compared with 5$\sigma$ during fits). }   
\label{fig:spec_phase}
\end{figure*}

The significance of the cyclotron line was estimated using the $F$-statistic \citep[e.g.,][]{orl12,sar15,bru18}. The \texttt{XSpec} script \texttt{fakeit} was used to simulate $2 \times 10^{6}$ spectra based on the model representing the null hypothesis, i.e., an absorbed, cutoff power law with an iron line but without a \texttt{cyclabs} component (``null model'' or M2). Each simulated spectrum was binned in the same way as the observed dataset, fit with the null model and its $\chi^{2}$ was recorded. Then, each simulated spectrum was fit with the ``best-fitting model'' (M3), which is the null model plus a \texttt{cyclabs} component whose parameters were allowed to vary within the 90\%-confidence region of the best-fitting values, and its $\chi^{2}$ was noted. These simulations yielded a distribution of reduced (i.e., normalized by the d.o.f.) ratios $F_{\mathrm{stat}} = \chi_{0}^{2} / \chi_{1}^{2}$, where the subscripts 0 and 1 denote the null model (M2) and the best-fitting model (M3), respectively. Every ratio from the simulations was less than the observed $F_{\mathrm{stat}}=1.051$ with the largest simulated value being $F_{\mathrm{stat}}=1.033$. From this distribution, we infer that the cyclotron line was significant at a level of at least 5$\sigma$ after accounting for the number of trials.

\begin{deluxetable*}{l | cc | cc | cc |}
\tablecaption{Fitting parameters of the phase-resolved non-flare spectra of IGR J16320$-$4751 with three empirical models.   }
\tablehead{
\colhead{} & \colhead{Valley} & \colhead{Peak} \vline & \colhead{Valley} &  \colhead{Peak} \vline & \colhead{Valley} &  \colhead{Peak} \\ 
\cline{2-7}
\colhead{} & \multicolumn{2}{c}{Model 1}  & \multicolumn{2}{|c|}{Model 2} & \multicolumn{2}{c}{Model 3} 
} 

\startdata       
      &     \multicolumn{6}{c|}{photoelectric absorption ({\tt Tbabs}) }     
\\   
$N_{\mathrm{H}}^{a}$
& 18.2$\pm$1.2 
& 20.9$\pm$1.4 
& 13.3$_{-1.5}^{+1.6}$ 
& 13.5$_{-1.9}^{+2.0}$ 
& 10.5$\pm$1.9 
& 8.4$_{-3.5}^{+3.3}$  
\\  
      &     \multicolumn{6}{c|}{cutoff power law ({\tt CutoffPL})}     
\\
$\Gamma$ 
& 0.37$\pm$0.05 
& 0.60$\pm$0.06 
& 0.13$\pm$0.08   
& 0.22$\pm$0.11 
& $-$0.10$\pm$0.12 
& $-0.25_{-0.28}^{+0.21}$  
\\
$E_{\mathrm{cut}}^{b}$   
& 14.6$_{-0.6}^{+0.7}$ 
& 14.9$\pm$0.8 
& 12.7$_{-0.6}^{+0.7}$   
& 11.9$_{-0.7}^{+0.8}$ 
& 11.2$\pm$0.7 
& 9.6$_{-1.0}^{+0.9}$ 
\\
$f_{\mathrm{cut}}^{c}$   
& 25.9$_{-2.6}^{+2.9}$ 
& 49.5$_{-5.6}^{+6.5}$ 
& 15.8$_{-2.2}^{+2.8}$   
& 22.2$_{-4.4}^{+5.2}$ 
& 10.4$_{-2.1}^{+2.6}$ 
& 9.8$_{-4.0}^{+4.7}$  
\\
      &     \multicolumn{6}{c|}{Fe K$_{\alpha}$ line ({\tt Gauss})}     
\\
$E_{\mathrm{Fe}}^{b}$   
& \nodata  
& \nodata  
& 6.34$\pm$0.07
& 6.31$\pm$0.11  
& 6.30$\pm$0.07 
& 6.14$_{-0.14}^{+0.32}$
\\
$\sigma_{\mathrm{Fe}}^{b}$   
& \nodata  
& \nodata  
& 0.34$_{-0.14}^{+0.10}$  
& 0.64$_{-0.15}^{+0.16}$ 
& 0.40$_{-0.10}^{+0.09}$ 
& 0.74$_{-0.30}^{+0.18}$ 
\\
$f_{\mathrm{Fe}}^{c}$   
& \nodata  
& \nodata  
& 1.45$_{-0.38}^{+0.36}$   
& 3.11$_{-0.76}^{+0.89}$ 
& 1.77$_{-0.38}^{+0.39}$ 
& 4.17$_{-1.87}^{+1.63}$ 
\\
      &     \multicolumn{6}{c|}{CRSF (\texttt{cyclabs})}     
\\
$E_{\mathrm{CRSF}}$ $^{b}$   
& \nodata  
& \nodata  
& \nodata 
& \nodata  
& 15.1$\pm$1.0  
& 13.8$_{-3.3}^{+2.8}$ 
\\
$\sigma_{\mathrm{CRSF}}^{b}$   
& \nodata  
& \nodata  
& \nodata   
& \nodata  
& 3.9$_{-1.4}^{+1.7}$
& 5.8$_{-3.2}^{+3.9}$ 
\\
$\tau_{\mathrm{CRSF}}^{d}$   
& \nodata  
& \nodata  
& \nodata 
& \nodata  
& 0.10$\pm$0.03
& 0.15$_{-0.05}^{+0.11}$ 
\\
\cline{1-7}
$\chi_{red}^{2}$ (d.o.f.)  
& 1.15 (1197) 
& 1.25 (1029) 
& 1.07 (1194) 
& 1.16 (1026) 
& 1.04 (1191)  
& 1.13 (1023)  
\\   
$F^{e}$ 
& 8.25$_{-0.10}^{+0.06}$
& 9.67$_{-0.16}^{+0.08}$ 
& 8.29$_{-0.14}^{+0.06}$  
& 9.75$_{-0.31}^{+0.07}$ 
& 8.35$_{-0.36}^{+0.05}$ 
& 9.85$_{-0.50}^{+0.04}$
\\
\enddata
\tablecomments{ (a) equivalent hydrogen column density ($\times 10^{22}$ cm$^{-2}$); (b) in keV; (c) normalization at 1 keV ($\times10^{-4}$ ph s$^{-1}$ cm$^{-2}$ keV$^{-1}$); (d) optical depth; (e) model-derived flux in the 2--10 keV band ($\times10^{-11}$ ergs s$^{-1}$ cm$^{-2}$) } 
\label{table_PR_fit}
\end{deluxetable*}

\subsubsection{Phase-resolved Spectroscopy}

We performed phase-resolved spectroscopy focusing only on data from the non-flare epoch. The pulse profile was split according to phases belonging to the ``Valley'' (phases: 0--0.2; 0.6--1.0) and ``Peak'' (phases: 0.2--0.6) as shown in Fig. \ref{fig:spin_profile}. This allocated a total of 105,410$\pm$325 net counts to the Valley in 28.85 ks of effective exposure time, and 75,516$\pm$557 net counts to the Peak in 18.78 ks of effective exposure time, when summing the counts from both modules. 

\begin{figure}[]
\begin{center}
\includegraphics[width=0.45\textwidth]{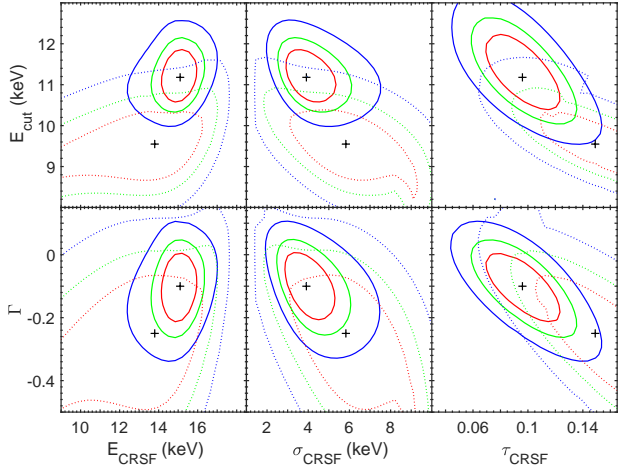}
\end{center}
 \caption{Same as Fig.\,\ref{fig:steppar}, with identical limits for the axes, for phases corresponding to the Peak (dotted lines) and the Valley (solid lines).}   
\label{fig:steppar_phase}
\end{figure}

Figure \ref{fig:spec_phase} presents the phase-resolved spectra fit with the three models introduced earlier, and Table \ref{table_PR_fit} lists the model parameters. Model 3 continued to provide the best fit. Count rates and model-derived fluxes were between 10\% and 20\% higher during the Peak than they were during the Valley, which is consistent with the pulsed fraction. Figure\,\ref{fig:steppar_phase} shows confidence regions for the continuum and CRSF parameters during the Peak and Valley phases.

During the Valley phase, there were negative residuals in the spectral fits for the 30--35-keV energy range. This energy is a little higher than would be expected for the harmonic to a candidate cyclotron line at 15 keV. Introducing a second cyclotron component did not reduce the $\chi^{2}$, and the component's parameters could not be constrained without holding others constant. This dip was not observed in the spectral residuals during the Peak phase.

\section{DISCUSSION} \label{sec:discussion}

The unusual properties of obscured, slowly-rotating pulsars such as \igr\ trace their origin to the interaction between the NS's magnetosphere and the inhomogeneous accretion stream from the stellar wind \citep[e.g.,][]{gre07,pat07,boz08,osk12,man12,hai20}. The magnetic field strength ($B$) of the NS can be directly measured by observing Cyclotron Resonance Scattering Features (CRSFs) which are generally observed as absorption lines between 10 and 100 keV. They arise through resonant scattering of photons emitted by electrons moving perpendicular to the $B$-field and whose energies ($E$) are discretized into integer multiples of the fundamental Landau level: $E_{\mathrm{cyc}} \sim 12 (B/ 10^{12}$ G) keV \citep{tru78,cob02}. 

Thanks to \nustar, we were able to perform a spectroscopic study of \igr\ with unprecedented energy resolution and sensitivity above 10 keV. A spectral model consisting of an absorbed power law left residuals near the known iron K$\alpha$ line energy of 6.4 keV. However, there were also residuals around 14 keV due to a possible CRSF. The addition of a cyclotron line to the model improved the fit quality enough that the distribution of $F$-statistics implied a detection significance of 5$\sigma$. The energy of the candidate cyclotron line was close to, but not statistically compatible with, the cutoff energy. Plus, in every instance of our simulated spectra, a model that included a cutoff and a cyclotron led to a better fit than a model with a cutoff alone. One alternative is that the candidate cyclotron line is an example of the ``10-keV bump'' noted in other accreting X-ray pulsars \citep{cob02,fer09}. However, a model in which the cyclotron line is replaced with with a Gaussian emission line or a Compton hump both led to a poorer quality fit. As an additional test, we followed the procedure in \citet{bot22} where the cyclotron energy was stepped in increments of 0.4 keV (\nustar's energy resolution) through the full 3--79 keV band, until the best fit was obtained based on $\chi^{2}$. Once again, the cyclotron was detected significantly around 14 keV (according to the reduced $F$-statistic distribution) and away from the cutoff energy and the possible bump. A cyclotron line energy of 14 keV corresponds to a $B$-field magnitude of $1.2 \times 10^{12}$ G, neglecting the gravitational redshift of the emission region. This is the first time that the magnetic field of this source has been measured. It is not particularly strong compared with its peers \citep{sta19}.

Cyclotron lines show variability with luminosity and pulse phase, and are often most significantly detected during certain phases \citep[e.g.,][]{suc12}. The candidate cyclotron line in \igr\ was easier to detect during the non-flare epoch, and during the Valley when analyzing by phase. We saw no evidence of an increase in the line energy with pulse phase, as was seen in Her X-1 \citep[e.g.,][]{soo90,vas13}, nor were there significant negative/positive correlations with luminosity \citep[][and references therein]{sta19}. 

\begin{figure}[]
\begin{center}
\includegraphics[width=0.45\textwidth]{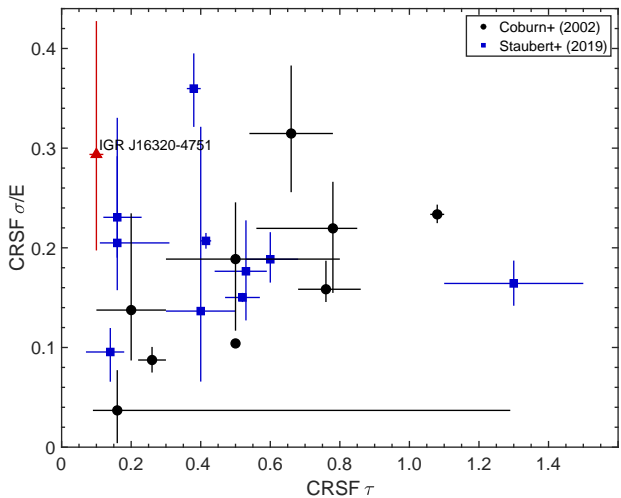}
\end{center}
\caption{Relative width ($\sigma/E$) versus optical depth ($\tau$) of the CRSF feature \citep[adapted from ][]{cob02}. The location of \igr\ in the parameter space is indicated by the red triangle, while other HMXBs appear as black circles from \citet{cob02} and blue squares from \citet{sta19}, and references therein.}
\label{fig:CRSF_W-OpD}
\end{figure}

The cyclotron scattering cross-section, and by extension the optical depth, are strongly affected by the viewing angle relative to the axis defined by the magnetic field \citep[e.g.,][]{sch17a,sch17b}. The optical depth of $\tau_{\mathrm{cyc}} = 0.1$ in \igr\ was lower than those of ten other sources reported in \citet{cob02}, which ranged from 0.16 to 2.1. Also, the ratio of width to energy ($\sigma_{\mathrm{cyc}}$/$E_{\mathrm{cyc}} = 0.3$) that we measured was more than twice as large as expected based on the trend found in \citet{cob02} where deeper CRSFs tended to be broader (Fig. \ref{fig:CRSF_W-OpD}). On the other hand, our values for \igr\ occupy a region of the parameter space that was inaccessible to the systematic study of \rxte\ data by \citet{cob02}, but that can now be explored by \nustar, which we did by adding 12 more HMXBs listed in \citet{sta19}, and references therein. Where \igr\ was once an outlier compared with the HMXBs of \cite{cob02}, it now has company in that it overlaps statistically with \object{IGR J18027$-$2016} and \object{KS~1947+300}. Both of them had many years pass between their discoveries \citep{aug03,bor90} and the detection of weak cyclotron lines by \nustar\ \citep{lut17,fur14b}. 

A coherent modulation with a period of 1,308.8$\pm$0.4 s was measured by \nustar. This is known to be the spin period of the NS in \igr. \citet{rod06} found an average value of 1,303 s between two separate \xmm\ observations from 2004, while \citet{gar18} report a period of $\sim$1,300 s from \xmm\ observations between 2003 and 2008. Assuming there was no torque reversal \citep{bil97}, a 5.8-s difference between measurements from 2004 and 2015 (3,943 d) indicates a spin-down trend with a period derivative $\dot{P} \sim 2 \times 10^{-8}$ s s$^{-1}$. Attributing the slowing down of the NS to magnetic braking or propellor effects \citep{ill75} would require a higher magnetic field strength and pulsation frequency than what we measured for \igr. The accretion of material with negative angular momentum appears to be the most likely explanation, and this could proceed either through an inhomogeneous stellar wind \citep{sha12}, or via a short-lived accretion disk. In addition to provoking outbursts in the SFXT \object{IGR~J17544$-$2619} \citep{rom15}, transient accretion disks may explain long-term changes in the spin period of other HMXBs with supergiant stars such as \object{OAO~1657$-$415} \citep{jen12} and \object{IGR~J16393$-$4643} \citep{bod16}. The spin period derivatives of \igr\ and \object{IGR~J16393$-$4643} are equal in magnitude, but unequal in direction: the former is slowing down while the latter is speeding up.

The pulsation was detected in every energy band that we analyzed except for 25--79 keV where few counts remained since the spectral continuum decays exponentially $\gtrsim$10 keV. The pulsed fraction in 3--6 keV was twice as large as that of 12--25 keV, which is surprising given that the pulsed fraction stayed constant with energy in 2004. It is also surprising given that in accreting X-ray pulsars, the pulsed fraction tends to increase with energy \citep{nag89,bil97,mus22}. This suggests a physical change in this system in the intervening 11 years which could include, for example, a reconfiguration of the magnetic field, a change in the energy dependence over the beam, or a change in beam pattern or size. The low significance and low pulsed fraction of the modulation at energies above 10 keV could explain the weakness of the candidate cyclotron line. 

\igr\ is a persistent X-ray source with a count rate that stayed relatively constant over almost 2 decades of \swift/BAT monitoring in 15--50 keV \citep{kri13}. Still, the source underwent small flares with the largest flare reaching a count rate a factor 20 times the average. An orbital modulation with a refined period of 8.9912$\pm$0.0078 d was found, in agreement with previous measurements \citep{cor05,lev11,gar18}. The \nustar\ observation covered orbital phases 0.97--0.11 where phase 0 ($=1$) represents the minimum point of the orbital profile. By extension, the \nustar\ observation occurred close to superior conjunction, i.e., when the X-ray emitting NS was furthest in its orbit with respect to the observer. With an eccentricity of 0.2 and an inclination angle of 62$^{\circ}$ \citep{gar18}, the source is not eclipsing so attenuation of X-rays at this phase is likely due to absorption by the companion star's wind. The \xmm\ observation of \citet{rod06} coincided with phases 0.40--0.46, i.e., just before the maximum point of the orbital profile, or near inferior conjunction. 

The unabsorbed 2--10-keV flux reported by \citet{rod06} during their non-flare epoch was $9.2\times10^{-11}$ \ergscm, which is equivalent to the absorption-corrected flux for the non-flare epoch that we found with \nustar: $1.1\times10^{-10}$ \ergscm. This is somewhat surprising given that the 2004 \xmm\ and 2015 \nustar\ observations occurred, respectively, near the highest and lowest points in the orbital profile. However, \citet{gar18} reported on a 2008 \xmm\ observation taken during phase 0 where the flux was $2.17\times10^{-10}$ \ergscm\ (0.15--12 keV), so the flux discrepancy is probably due to the stochastic variability of the source. 

Based on infrared spectroscopy of the counterpart to \igr, \citet{rah08} estimated a source distance of 3.5 kpc. At this distance, the intrinsic (i.e., absorption-corrected) 2--10-keV source luminosity would be $1.7 \times 10^{35}$ \ergs\ outside the flare, and $5.1 \times 10^{35}$ \ergs\ during the flare. No objects listed in the \emph{Gaia} DR3 \citep{gai16,gai21} catalog of parallax-derived distances \citep{bai21} were within 5$^{\prime\prime}$ of the 4XMM position of \igr. If we assume a distance of 10 kpc instead, then the luminosities during the non-flare and flare epochs would be $1.4 \times 10^{36}$ \ergs\ and $4.2 \times 10^{36}$ \ergs, respectively.

\section{CONCLUSIONS} \label{sec:summary}

\nustar\ gave an exclusive look at the accreting X-ray pulsar \igr\ in an energy band above 10 keV that has not been covered with as fine a spectral resolution with other telescopes. The spectrum of \igr\ was best fit by introducing a CRSF at $\sim$14 keV in addition to a cutoff power-law continuum and an Fe K$\alpha$ line. If confirmed, the cyclotron line would represent the first direct measurement of a $1.2 \times 10^{12}$-G magnetic field for the neutron star in \igr. In this 2015 study, the pulsed fraction showed a significant negative correlation with energy, whereas the pulsed fraction remained constant in 2004. This suggests that the system's magnetically-driven accretion geometry changed between observations.

\nustar\ provided new insights into the evolution of the line properties on long and short time scales in HMXBs known to have CRSFs, e.g., \object{Her~X-1} and \object{Vela~X-1} \citep{fur13,fur14a}. \nustar\ has also uncovered cyclotron lines for HMXBs not previously known to have them \citep[e.g.,][]{fur14b,shr14,bha15,bod16}. With $\sim$35 known CRSF sources, including candidate cyclotron sources of which \igr\ is now a member, the increasing sample size will permit studies of the relationship between the $B$-field and other properties, such as luminosity, companion type, spin period, and orbital period \citep[e.g.,][]{sch14,sta19,chr19}.

\begin{acknowledgments}

The authors thank the anonymous Referee whose constructive criticism helped improve our discussion of the cyclotron line. JAT and AC acknowledge partial support from NASA grant NNX15AV22G under the \nustar\ Guest Observer program. MC acknowledges financial support from the Centre National d’Etudes Spatiales (CNES). PR acknowledges financial contribution from the agreements ASI-INAF I/037/12/0 and ASI-INAF n. 2017-14-H.0. The scientific results reported in this article are based on data from the \nustar\ mission, a project led by the California Institute of Technology, managed by the Jet Propulsion Laboratory, and funded by the National Aeronautics and Space Administration. This research has made use of: the \nustar\ Data Analysis Software (NuSTARDAS) jointly developed by the ASI Science Data Center (ASDC, Italy) and the California Institute of Technology; data obtained from the High Energy Astrophysics Science Archive Research Center (HEASARC) provided by NASA's Goddard Space Flight Center; NASA's Astrophysics Data System Bibliographic Services; and the SIMBAD database operated at CDS, Strasbourg, France.

 \end{acknowledgments}

%

\vspace{5mm}
\facilities{\nustar, \swift\ (XRT and BAT)}


\software{HEASoft v6.29 \citep{hea14},
		FTOOLS v6.29 \citep{bla99},
		NuSTARDAS v2.1.1,
		XSpec v12.12.0 \citep{arn96},
		Matlab R2021b \citep{mat21}}


\bibliography{NuSTAR_16320}{}

\begin{thebibliography}{}
\expandafter\ifx\csname natexlab\endcsname\relax\def\natexlab#1{#1}\fi
\providecommand{\url}[1]{\href{#1}{#1}}
\providecommand{\dodoi}[1]{doi:~\href{http://doi.org/#1}{\nolinkurl{#1}}}
\providecommand{\doeprint}[1]{\href{http://ascl.net/#1}{\nolinkurl{http://ascl.net/#1}}}
\providecommand{\doarXiv}[1]{\href{https://arxiv.org/abs/#1}{\nolinkurl{https://arxiv.org/abs/#1}}}

\bibitem[{{Abdollahi} {et~al.}(2020){Abdollahi}, {Acero}, {Ackermann},
  {Ajello}, {Atwood}, {Axelsson}, {Baldini}, {Ballet}, {Barbiellini},
  {Bastieri}, {Becerra Gonzalez}, {Bellazzini}, {Berretta}, {Bissaldi},
  {Blandford}, {Bloom}, {Bonino}, {Bottacini}, {Brandt}, {Bregeon}, {Bruel},
  {Buehler}, {Burnett}, {Buson}, {Cameron}, {Caputo}, {Caraveo}, {Casandjian},
  {Castro}, {Cavazzuti}, {Charles}, {Chaty}, {Chen}, {Cheung}, {Chiaro},
  {Ciprini}, {Cohen-Tanugi}, {Cominsky}, {Coronado-Bl{\'a}zquez}, {Costantin},
  {Cuoco}, {Cutini}, {D'Ammando}, {DeKlotz}, {de la Torre Luque}, {de Palma},
  {Desai}, {Digel}, {Di Lalla}, {Di Mauro}, {Di Venere}, {Dom{\'\i}nguez},
  {Dumora}, {Fana Dirirsa}, {Fegan}, {Ferrara}, {Franckowiak}, {Fukazawa},
  {Funk}, {Fusco}, {Gargano}, {Gasparrini}, {Giglietto}, {Giommi}, {Giordano},
  {Giroletti}, {Glanzman}, {Green}, {Grenier}, {Griffin}, {Grondin}, {Grove},
  {Guiriec}, {Harding}, {Hayashi}, {Hays}, {Hewitt}, {Horan},
  {J{\'o}hannesson}, {Johnson}, {Kamae}, {Kerr}, {Kocevski}, {Kovac'evic'},
  {Kuss}, {Landriu}, {Larsson}, {Latronico}, {Lemoine-Goumard}, {Li},
  {Liodakis}, {Longo}, {Loparco}, {Lott}, {Lovellette}, {Lubrano}, {Madejski},
  {Maldera}, {Malyshev}, {Manfreda}, {Marchesini}, {Marcotulli},
  {Mart{\'\i}-Devesa}, {Martin}, {Massaro}, {Mazziotta}, {McEnery}, {Mereu},
  {Meyer}, {Michelson}, {Mirabal}, {Mizuno}, {Monzani}, {Morselli},
  {Moskalenko}, {Negro}, {Nuss}, {Ojha}, {Omodei}, {Orienti}, {Orlando},
  {Ormes}, {Palatiello}, {Paliya}, {Paneque}, {Pei}, {Pe{\~n}a-Herazo},
  {Perkins}, {Persic}, {Pesce-Rollins}, {Petrosian}, {Petrov}, {Piron}, {Poon},
  {Porter}, {Principe}, {Rain{\`o}}, {Rando}, {Razzano}, {Razzaque}, {Reimer},
  {Reimer}, {Remy}, {Reposeur}, {Romani}, {Saz Parkinson}, {Schinzel},
  {Serini}, {Sgr{\`o}}, {Siskind}, {Smith}, {Spandre}, {Spinelli}, {Strong},
  {Suson}, {Tajima}, {Takahashi}, {Tak}, {Thayer}, {Thompson}, {Tibaldo},
  {Torres}, {Torresi}, {Valverde}, {Van Klaveren}, {van Zyl}, {Wood},
  {Yassine}, \& {Zaharijas}}]{abd20}
{Abdollahi}, S., {Acero}, F., {Ackermann}, M., {et~al.} 2020, \apjs, 247, 33,
  \dodoi{10.3847/1538-4365/ab6bcb}

\bibitem[{{Acero} {et~al.}(2015){Acero}, {Ackermann}, {Ajello}, {Albert},
  {Atwood}, {Axelsson}, {Baldini}, {Ballet}, {Barbiellini}, {Bastieri},
  {Belfiore}, {Bellazzini}, {Bissaldi}, {Blandford}, {Bloom}, {Bogart},
  {Bonino}, {Bottacini}, {Bregeon}, {Britto}, {Bruel}, {Buehler}, {Burnett},
  {Buson}, {Caliandro}, {Cameron}, {Caputo}, {Caragiulo}, {Caraveo},
  {Casandjian}, {Cavazzuti}, {Charles}, {Chaves}, {Chekhtman}, {Cheung},
  {Chiang}, {Chiaro}, {Ciprini}, {Claus}, {Cohen-Tanugi}, {Cominsky}, {Conrad},
  {Cutini}, {D'Ammando}, {de Angelis}, {DeKlotz}, {de Palma}, {Desiante},
  {Digel}, {Di Venere}, {Drell}, {Dubois}, {Dumora}, {Favuzzi}, {Fegan},
  {Ferrara}, {Finke}, {Franckowiak}, {Fukazawa}, {Funk}, {Fusco}, {Gargano},
  {Gasparrini}, {Giebels}, {Giglietto}, {Giommi}, {Giordano}, {Giroletti},
  {Glanzman}, {Godfrey}, {Grenier}, {Grondin}, {Grove}, {Guillemot}, {Guiriec},
  {Hadasch}, {Harding}, {Hays}, {Hewitt}, {Hill}, {Horan}, {Iafrate}, {Jogler},
  {J{\'o}hannesson}, {Johnson}, {Johnson}, {Johnson}, {Johnson}, {Kamae},
  {Kataoka}, {Katsuta}, {Kuss}, {La Mura}, {Landriu}, {Larsson}, {Latronico},
  {Lemoine-Goumard}, {Li}, {Li}, {Longo}, {Loparco}, {Lott}, {Lovellette},
  {Lubrano}, {Madejski}, {Massaro}, {Mayer}, {Mazziotta}, {McEnery},
  {Michelson}, {Mirabal}, {Mizuno}, {Moiseev}, {Mongelli}, {Monzani},
  {Morselli}, {Moskalenko}, {Murgia}, {Nuss}, {Ohno}, {Ohsugi}, {Omodei},
  {Orienti}, {Orlando}, {Ormes}, {Paneque}, {Panetta}, {Perkins},
  {Pesce-Rollins}, {Piron}, {Pivato}, {Porter}, {Racusin}, {Rando}, {Razzano},
  {Razzaque}, {Reimer}, {Reimer}, {Reposeur}, {Rochester}, {Romani},
  {Salvetti}, {S{\'a}nchez-Conde}, {Saz Parkinson}, {Schulz}, {Siskind},
  {Smith}, {Spada}, {Spandre}, {Spinelli}, {Stephens}, {Strong}, {Suson},
  {Takahashi}, {Takahashi}, {Tanaka}, {Thayer}, {Thayer}, {Thompson},
  {Tibaldo}, {Tibolla}, {Torres}, {Torresi}, {Tosti}, {Troja}, {Van Klaveren},
  {Vianello}, {Winer}, {Wood}, {Wood}, {Zimmer}, \& {Fermi-LAT
  Collaboration}}]{ace15}
{Acero}, F., {Ackermann}, M., {Ajello}, M., {et~al.} 2015, \apjs, 218, 23,
  \dodoi{10.1088/0067-0049/218/2/23}

\bibitem[{{Aharonian} {et~al.}(2006){Aharonian}, {Akhperjanian}, {Bazer-Bachi},
  {Beilicke}, {Benbow}, {Berge}, {Bernl{\"o}hr}, {Boisson}, {Bolz}, {Borrel},
  {Braun}, {Breitling}, {Brown}, {Chadwick}, {Chounet}, {Cornils},
  {Costamante}, {Degrange}, {Dickinson}, {Djannati-Ata{\"\i}}, {Drury},
  {Dubus}, {Emmanoulopoulos}, {Espigat}, {Feinstein}, {Fontaine}, {Fuchs},
  {Funk}, {Gallant}, {Giebels}, {Gillessen}, {Glicenstein}, {Goret},
  {Hadjichristidis}, {Hauser}, {Heinzelmann}, {Henri}, {Hermann}, {Hinton},
  {Hofmann}, {Holleran}, {Horns}, {Jacholkowska}, {de Jager}, {Kh{\'e}lifi},
  {Komin}, {Konopelko}, {Latham}, {Le Gallou}, {Lemi{\`e}re},
  {Lemoine-Goumard}, {Leroy}, {Lohse}, {Martin}, {Martineau-Huynh},
  {Marcowith}, {Masterson}, {McComb}, {de Naurois}, {Nolan}, {Noutsos},
  {Orford}, {Osborne}, {Ouchrif}, {Panter}, {Pelletier}, {Pita},
  {P{\"u}hlhofer}, {Punch}, {Raubenheimer}, {Raue}, {Raux}, {Rayner}, {Reimer},
  {Reimer}, {Ripken}, {Rob}, {Rolland}, {Rowell}, {Sahakian}, {Saug{\'e}},
  {Schlenker}, {Schlickeiser}, {Schuster}, {Schwanke}, {Siewert}, {Sol},
  {Spangler}, {Steenkamp}, {Stegmann}, {Tavernet}, {Terrier}, {Th{\'e}oret},
  {Tluczykont}, {Vasileiadis}, {Venter}, {Vincent}, {V{\"o}lk}, \&
  {Wagner}}]{aha06}
{Aharonian}, F., {Akhperjanian}, A.~G., {Bazer-Bachi}, A.~R., {et~al.} 2006,
  \apj, 636, 777, \dodoi{10.1086/498013}

\bibitem[{{Arnaud}(1996)}]{arn96}
{Arnaud}, K.~A. 1996, in Astronomical Society of the Pacific Conference Series,
  Vol. 101, Astronomical Data Analysis Software and Systems V, ed. G.~H.
  {Jacoby} \& J.~{Barnes}, 17

\bibitem[{{Augello} {et~al.}(2003){Augello}, {Iaria}, {Robba}, {Di Salvo},
  {Burderi}, {Lavagetto}, \& {Stella}}]{aug03}
{Augello}, G., {Iaria}, R., {Robba}, N.~R., {et~al.} 2003, \apjl, 596, L63,
  \dodoi{10.1086/379092}

\bibitem[{{Bailer-Jones} {et~al.}(2021){Bailer-Jones}, {Rybizki}, {Fouesneau},
  {Demleitner}, \& {Andrae}}]{bai21}
{Bailer-Jones}, C.~A.~L., {Rybizki}, J., {Fouesneau}, M., {Demleitner}, M., \&
  {Andrae}, R. 2021, \aj, 161, 147, \dodoi{10.3847/1538-3881/abd806}

\bibitem[{{Ballhausen} {et~al.}(2017){Ballhausen}, {Pottschmidt}, {F{\"u}rst},
  {Wilms}, {Tomsick}, {Schwarm}, {Stern}, {Kretschmar}, {Caballero},
  {Harrison}, {Boggs}, {Christensen}, {Craig}, {Hailey}, \& {Zhang}}]{bal17}
{Ballhausen}, R., {Pottschmidt}, K., {F{\"u}rst}, F., {et~al.} 2017, \aap, 608,
  A105, \dodoi{10.1051/0004-6361/201730845}

\bibitem[{{Bellm} {et~al.}(2014){Bellm}, {F{\"u}rst}, {Pottschmidt}, {Tomsick},
  {Boggs}, {Chakrabarty}, {Christensen}, {Craig}, {Hailey}, {Harrison},
  {Stern}, {Walton}, {Wilms}, \& {Zhang}}]{bel14}
{Bellm}, E.~C., {F{\"u}rst}, F., {Pottschmidt}, K., {et~al.} 2014, \apj, 792,
  108, \dodoi{10.1088/0004-637X/792/2/108}

\bibitem[{{Bhalerao} {et~al.}(2015){Bhalerao}, {Romano}, {Tomsick},
  {Natalucci}, {Smith}, {Bellm}, {Boggs}, {Chakrabarty}, {Christensen},
  {Craig}, {Fuerst}, {Hailey}, {Harrison}, {Krivonos}, {Lu}, {Madsen}, {Stern},
  {Younes}, \& {Zhang}}]{bha15}
{Bhalerao}, V., {Romano}, P., {Tomsick}, J., {et~al.} 2015, \mnras, 447, 2274,
  \dodoi{10.1093/mnras/stu2495}

\bibitem[{{Bildsten} {et~al.}(1997){Bildsten}, {Chakrabarty}, {Chiu}, {Finger},
  {Koh}, {Nelson}, {Prince}, {Rubin}, {Scott}, {Stollberg}, {Vaughan},
  {Wilson}, \& {Wilson}}]{bil97}
{Bildsten}, L., {Chakrabarty}, D., {Chiu}, J., {et~al.} 1997, \apjs, 113, 367,
  \dodoi{10.1086/313060}

\bibitem[{{Blackburn} {et~al.}(1999){Blackburn}, {Shaw}, {Payne}, {Hayes}, \&
  {Heasarc}}]{bla99}
{Blackburn}, J.~K., {Shaw}, R.~A., {Payne}, H.~E., {Hayes}, J.~J.~E., \&
  {Heasarc}. 1999, {FTOOLS: A general package of software to manipulate FITS
  files}, Astrophysics Source Code Library, record ascl:9912.002.
\newblock \doeprint{9912.002}

\bibitem[{{Bodaghee} {et~al.}(2006){Bodaghee}, {Walter}, {Zurita Heras},
  {Bird}, {Courvoisier}, {Malizia}, {Terrier}, \& {Ubertini}}]{bod06}
{Bodaghee}, A., {Walter}, R., {Zurita Heras}, J.~A., {et~al.} 2006, \aap, 447,
  1027, \dodoi{10.1051/0004-6361:20053809}

\bibitem[{{Bodaghee} {et~al.}(2016){Bodaghee}, {Tomsick}, {Fornasini},
  {Krivonos}, {Stern}, {Mori}, {Rahoui}, {Boggs}, {Christensen}, {Craig},
  {Hailey}, {Harrison}, \& {Zhang}}]{bod16}
{Bodaghee}, A., {Tomsick}, J.~A., {Fornasini}, F.~M., {et~al.} 2016, \apj, 823,
  146, \dodoi{10.3847/0004-637X/823/2/146}

\bibitem[{{Borozdin} {et~al.}(1990){Borozdin}, {Gilfanov}, {Sunyaev},
  {Churazov}, {Loznikov}, {Yamburenko}, {Skinner}, {Patterson}, {Willmore},
  {Emam}, {Brinkman}, {Heise}, {Int-Zand}, \& {Jager}}]{bor90}
{Borozdin}, K., {Gilfanov}, M., {Sunyaev}, R., {et~al.} 1990, Soviet Astronomy
  Letters, 16, 345

\bibitem[{{Bottacini}(2022)}]{bot22}
{Bottacini}, E. 2022, \mnras, 515, 3174, \dodoi{10.1093/mnras/stac1890}

\bibitem[{{Bozzo} {et~al.}(2008){Bozzo}, {Falanga}, \& {Stella}}]{boz08}
{Bozzo}, E., {Falanga}, M., \& {Stella}, L. 2008, \apj, 683, 1031,
  \dodoi{10.1086/589990}

\bibitem[{{Bozzo} {et~al.}(2015){Bozzo}, {Romano}, {Ducci}, {Bernardini}, \&
  {Falanga}}]{boz15}
{Bozzo}, E., {Romano}, P., {Ducci}, L., {Bernardini}, F., \& {Falanga}, M.
  2015, Advances in Space Research, 55, 1255, \dodoi{10.1016/j.asr.2014.11.012}

\bibitem[{{Brumback} {et~al.}(2018){Brumback}, {Hickox}, {F{\"u}rst},
  {Pottschmidt}, {Hemphill}, {Tomsick}, {Wilms}, \& {Ballhausen}}]{bru18}
{Brumback}, M.~C., {Hickox}, R.~C., {F{\"u}rst}, F.~S., {et~al.} 2018, \apj,
  852, 132, \dodoi{10.3847/1538-4357/aa9e91}

\bibitem[{{Cash}(1979)}]{cas79}
{Cash}, W. 1979, \apj, 228, 939, \dodoi{10.1086/156922}

\bibitem[{{Christodoulou} {et~al.}(2019){Christodoulou}, {Laycock}, \&
  {Kazanas}}]{chr19}
{Christodoulou}, D.~M., {Laycock}, S. G.~T., \& {Kazanas}, D. 2019, Research in
  Astronomy and Astrophysics, 19, 146, \dodoi{10.1088/1674-4527/19/10/146}

\bibitem[{{Coburn} {et~al.}(2002){Coburn}, {Heindl}, {Rothschild}, {Gruber},
  {Kreykenbohm}, {Wilms}, {Kretschmar}, \& {Staubert}}]{cob02}
{Coburn}, W., {Heindl}, W.~A., {Rothschild}, R.~E., {et~al.} 2002, \apj, 580,
  394, \dodoi{10.1086/343033}

\bibitem[{{Coleiro} {et~al.}(2013){Coleiro}, {Chaty}, {Zurita Heras}, {Rahoui},
  \& {Tomsick}}]{col13}
{Coleiro}, A., {Chaty}, S., {Zurita Heras}, J.~A., {Rahoui}, F., \& {Tomsick},
  J.~A. 2013, \aap, 560, A108, \dodoi{10.1051/0004-6361/201322382}

\bibitem[{{Corbet} {et~al.}(2005){Corbet}, {Barbier}, {Barthelmy}, {Cummings},
  {Fenimore}, {Gehrels}, {Hullinger}, {Krimm}, {Markwardt}, {Palmer},
  {Parsons}, {Sakamoto}, {Sato}, {Tueller}, \& {Swift-Survey Team}}]{cor05}
{Corbet}, R., {Barbier}, L., {Barthelmy}, S., {et~al.} 2005, The Astronomer's
  Telegram, 649, 1

\bibitem[{{Ferrigno} {et~al.}(2009){Ferrigno}, {Becker}, {Segreto}, {Mineo}, \&
  {Santangelo}}]{fer09}
{Ferrigno}, C., {Becker}, P.~A., {Segreto}, A., {Mineo}, T., \& {Santangelo},
  A. 2009, \aap, 498, 825, \dodoi{10.1051/0004-6361/200809373}

\bibitem[{{F{\"u}rst} {et~al.}(2013){F{\"u}rst}, {Grefenstette}, {Staubert},
  {Tomsick}, {Bachetti}, {Barret}, {Bellm}, {Boggs}, {Chenevez}, {Christensen},
  {Craig}, {Hailey}, {Harrison}, {Klochkov}, {Madsen}, {Pottschmidt}, {Stern},
  {Walton}, {Wilms}, \& {Zhang}}]{fur13}
{F{\"u}rst}, F., {Grefenstette}, B.~W., {Staubert}, R., {et~al.} 2013, \apj,
  779, 69, \dodoi{10.1088/0004-637X/779/1/69}

\bibitem[{{F{\"u}rst} {et~al.}(2014{\natexlab{a}}){F{\"u}rst}, {Pottschmidt},
  {Wilms}, {Kennea}, {Bachetti}, {Bellm}, {Boggs}, {Chakrabarty},
  {Christensen}, {Craig}, {Hailey}, {Harrison}, {Stern}, {Tomsick}, {Walton},
  \& {Zhang}}]{fur14b}
{F{\"u}rst}, F., {Pottschmidt}, K., {Wilms}, J., {et~al.} 2014{\natexlab{a}},
  \apjl, 784, L40, \dodoi{10.1088/2041-8205/784/2/L40}

\bibitem[{{F{\"u}rst} {et~al.}(2014{\natexlab{b}}){F{\"u}rst}, {Pottschmidt},
  {Wilms}, {Tomsick}, {Bachetti}, {Boggs}, {Christensen}, {Craig},
  {Grefenstette}, {Hailey}, {Harrison}, {Madsen}, {Miller}, {Stern}, {Walton},
  \& {Zhang}}]{fur14a}
---. 2014{\natexlab{b}}, \apj, 780, 133, \dodoi{10.1088/0004-637X/780/2/133}

\bibitem[{{Gaia Collaboration} {et~al.}(2016){Gaia Collaboration}, {Prusti},
  {de Bruijne}, {Brown}, {Vallenari}, {Babusiaux}, {Bailer-Jones}, {Bastian},
  {Biermann}, {Evans}, {Eyer}, {Jansen}, {Jordi}, {Klioner}, {Lammers},
  {Lindegren}, {Luri}, {Mignard}, {Milligan}, {Panem}, {Poinsignon},
  {Pourbaix}, {Randich}, {Sarri}, {Sartoretti}, {Siddiqui}, {Soubiran},
  {Valette}, {van Leeuwen}, {Walton}, {Aerts}, {Arenou}, {Cropper}, {Drimmel},
  {H{\o}g}, {Katz}, {Lattanzi}, {O'Mullane}, {Grebel}, {Holland}, {Huc},
  {Passot}, {Bramante}, {Cacciari}, {Casta{\~n}eda}, {Chaoul}, {Cheek}, {De
  Angeli}, {Fabricius}, {Guerra}, {Hern{\'a}ndez}, {Jean-Antoine-Piccolo},
  {Masana}, {Messineo}, {Mowlavi}, {Nienartowicz}, {Ord{\'o}{\~n}ez-Blanco},
  {Panuzzo}, {Portell}, {Richards}, {Riello}, {Seabroke}, {Tanga},
  {Th{\'e}venin}, {Torra}, {Els}, {Gracia-Abril}, {Comoretto},
  {Garcia-Reinaldos}, {Lock}, {Mercier}, {Altmann}, {Andrae}, {Astraatmadja},
  {Bellas-Velidis}, {Benson}, {Berthier}, {Blomme}, {Busso}, {Carry},
  {Cellino}, {Clementini}, {Cowell}, {Creevey}, {Cuypers}, {Davidson}, {De
  Ridder}, {de Torres}, {Delchambre}, {Dell'Oro}, {Ducourant}, {Fr{\'e}mat},
  {Garc{\'\i}a-Torres}, {Gosset}, {Halbwachs}, {Hambly}, {Harrison}, {Hauser},
  {Hestroffer}, {Hodgkin}, {Huckle}, {Hutton}, {Jasniewicz}, {Jordan},
  {Kontizas}, {Korn}, {Lanzafame}, {Manteiga}, {Moitinho}, {Muinonen},
  {Osinde}, {Pancino}, {Pauwels}, {Petit}, {Recio-Blanco}, {Robin}, {Sarro},
  {Siopis}, {Smith}, {Smith}, {Sozzetti}, {Thuillot}, {van Reeven}, {Viala},
  {Abbas}, {Abreu Aramburu}, {Accart}, {Aguado}, {Allan}, {Allasia},
  {Altavilla}, {{\'A}lvarez}, {Alves}, {Anderson}, {Andrei}, {Anglada Varela},
  {Antiche}, {Antoja}, {Ant{\'o}n}, {Arcay}, {Atzei}, {Ayache}, {Bach},
  {Baker}, {Balaguer-N{\'u}{\~n}ez}, {Barache}, {Barata}, {Barbier}, {Barblan},
  {Baroni}, {Barrado y Navascu{\'e}s}, {Barros}, {Barstow}, {Becciani},
  {Bellazzini}, {Bellei}, {Bello Garc{\'\i}a}, {Belokurov}, {Bendjoya},
  {Berihuete}, {Bianchi}, {Bienaym{\'e}}, {Billebaud}, {Blagorodnova},
  {Blanco-Cuaresma}, {Boch}, {Bombrun}, {Borrachero}, {Bouquillon}, {Bourda},
  {Bouy}, {Bragaglia}, {Breddels}, {Brouillet}, {Br{\"u}semeister},
  {Bucciarelli}, {Budnik}, {Burgess}, {Burgon}, {Burlacu}, {Busonero}, {Buzzi},
  {Caffau}, {Cambras}, {Campbell}, {Cancelliere}, {Cantat-Gaudin}, {Carlucci},
  {Carrasco}, {Castellani}, {Charlot}, {Charnas}, {Charvet}, {Chassat},
  {Chiavassa}, {Clotet}, {Cocozza}, {Collins}, {Collins}, {Costigan}, {Crifo},
  {Cross}, {Crosta}, {Crowley}, {Dafonte}, {Damerdji}, {Dapergolas}, {David},
  {David}, {De Cat}, {de Felice}, {de Laverny}, {De Luise}, {De March}, {de
  Martino}, {de Souza}, {Debosscher}, {del Pozo}, {Delbo}, {Delgado},
  {Delgado}, {di Marco}, {Di Matteo}, {Diakite}, {Distefano}, {Dolding}, {Dos
  Anjos}, {Drazinos}, {Dur{\'a}n}, {Dzigan}, {Ecale}, {Edvardsson}, {Enke},
  {Erdmann}, {Escolar}, {Espina}, {Evans}, {Eynard Bontemps}, {Fabre},
  {Fabrizio}, {Faigler}, {Falc{\~a}o}, {Farr{\`a}s Casas}, {Faye}, {Federici},
  {Fedorets}, {Fern{\'a}ndez-Hern{\'a}ndez}, {Fernique}, {Fienga}, {Figueras},
  {Filippi}, {Findeisen}, {Fonti}, {Fouesneau}, {Fraile}, {Fraser}, {Fuchs},
  {Furnell}, {Gai}, {Galleti}, {Galluccio}, {Garabato}, {Garc{\'\i}a-Sedano},
  {Gar{\'e}}, {Garofalo}, {Garralda}, {Gavras}, {Gerssen}, {Geyer}, {Gilmore},
  {Girona}, {Giuffrida}, {Gomes}, {Gonz{\'a}lez-Marcos},
  {Gonz{\'a}lez-N{\'u}{\~n}ez}, {Gonz{\'a}lez-Vidal}, {Granvik}, {Guerrier},
  {Guillout}, {Guiraud}, {G{\'u}rpide}, {Guti{\'e}rrez-S{\'a}nchez}, {Guy},
  {Haigron}, {Hatzidimitriou}, {Haywood}, {Heiter}, {Helmi}, {Hobbs},
  {Hofmann}, {Holl}, {Holland}, {Hunt}, {Hypki}, {Icardi}, {Irwin}, {Jevardat
  de Fombelle}, {Jofr{\'e}}, {Jonker}, {Jorissen}, {Julbe}, {Karampelas},
  {Kochoska}, {Kohley}, {Kolenberg}, {Kontizas}, {Koposov}, {Kordopatis},
  {Koubsky}, {Kowalczyk}, {Krone-Martins}, {Kudryashova}, {Kull}, {Bachchan},
  {Lacoste-Seris}, {Lanza}, {Lavigne}, {Le Poncin-Lafitte}, {Lebreton},
  {Lebzelter}, {Leccia}, {Leclerc}, {Lecoeur-Taibi}, {Lemaitre}, {Lenhardt},
  {Leroux}, {Liao}, {Licata}, {Lindstr{\o}m}, {Lister}, {Livanou}, {Lobel},
  {L{\"o}ffler}, {L{\'o}pez}, {Lopez-Lozano}, {Lorenz}, {Loureiro},
  {MacDonald}, {Magalh{\~a}es Fernandes}, {Managau}, {Mann}, {Mantelet},
  {Marchal}, {Marchant}, {Marconi}, {Marie}, {Marinoni}, {Marrese},
  {Marschalk{\'o}}, {Marshall}, {Mart{\'\i}n-Fleitas}, {Martino}, {Mary},
  {Matijevi{\v{c}}}, {Mazeh}, {McMillan}, {Messina}, {Mestre}, {Michalik},
  {Millar}, {Miranda}, {Molina}, {Molinaro}, {Molinaro}, {Moln{\'a}r},
  {Moniez}, {Montegriffo}, {Monteiro}, {Mor}, {Mora}, {Morbidelli}, {Morel},
  {Morgenthaler}, {Morley}, {Morris}, {Mulone}, {Muraveva}, {Musella},
  {Narbonne}, {Nelemans}, {Nicastro}, {Noval}, {Ord{\'e}novic},
  {Ordieres-Mer{\'e}}, {Osborne}, {Pagani}, {Pagano}, {Pailler}, {Palacin},
  {Palaversa}, {Parsons}, {Paulsen}, {Pecoraro}, {Pedrosa}, {Pentik{\"a}inen},
  {Pereira}, {Pichon}, {Piersimoni}, {Pineau}, {Plachy}, {Plum}, {Poujoulet},
  {Pr{\v{s}}a}, {Pulone}, {Ragaini}, {Rago}, {Rambaux}, {Ramos-Lerate},
  {Ranalli}, {Rauw}, {Read}, {Regibo}, {Renk}, {Reyl{\'e}}, {Ribeiro},
  {Rimoldini}, {Ripepi}, {Riva}, {Rixon}, {Roelens}, {Romero-G{\'o}mez},
  {Rowell}, {Royer}, {Rudolph}, {Ruiz-Dern}, {Sadowski}, {Sagrist{\`a}
  Sell{\'e}s}, {Sahlmann}, {Salgado}, {Salguero}, {Sarasso}, {Savietto},
  {Schnorhk}, {Schultheis}, {Sciacca}, {Segol}, {Segovia}, {Segransan},
  {Serpell}, {Shih}, {Smareglia}, {Smart}, {Smith}, {Solano}, {Solitro},
  {Sordo}, {Soria Nieto}, {Souchay}, {Spagna}, {Spoto}, {Stampa}, {Steele},
  {Steidelm{\"u}ller}, {Stephenson}, {Stoev}, {Suess}, {S{\"u}veges}, {Surdej},
  {Szabados}, {Szegedi-Elek}, {Tapiador}, {Taris}, {Tauran}, {Taylor},
  {Teixeira}, {Terrett}, {Tingley}, {Trager}, {Turon}, {Ulla}, {Utrilla},
  {Valentini}, {van Elteren}, {Van Hemelryck}, {van Leeuwen}, {Varadi},
  {Vecchiato}, {Veljanoski}, {Via}, {Vicente}, {Vogt}, {Voss}, {Votruba},
  {Voutsinas}, {Walmsley}, {Weiler}, {Weingrill}, {Werner}, {Wevers},
  {Whitehead}, {Wyrzykowski}, {Yoldas}, {{\v{Z}}erjal}, {Zucker}, {Zurbach},
  {Zwitter}, {Alecu}, {Allen}, {Allende Prieto}, {Amorim},
  {Anglada-Escud{\'e}}, {Arsenijevic}, {Azaz}, {Balm}, {Beck}, {Bernstein},
  {Bigot}, {Bijaoui}, {Blasco}, {Bonfigli}, {Bono}, {Boudreault}, {Bressan},
  {Brown}, {Brunet}, {Bunclark}, {Buonanno}, {Butkevich}, {Carret}, {Carrion},
  {Chemin}, {Ch{\'e}reau}, {Corcione}, {Darmigny}, {de Boer}, {de Teodoro}, {de
  Zeeuw}, {Delle Luche}, {Domingues}, {Dubath}, {Fodor}, {Fr{\'e}zouls},
  {Fries}, {Fustes}, {Fyfe}, {Gallardo}, {Gallegos}, {Gardiol}, {Gebran},
  {Gomboc}, {G{\'o}mez}, {Grux}, {Gueguen}, {Heyrovsky}, {Hoar}, {Iannicola},
  {Isasi Parache}, {Janotto}, {Joliet}, {Jonckheere}, {Keil}, {Kim},
  {Klagyivik}, {Klar}, {Knude}, {Kochukhov}, {Kolka}, {Kos}, {Kutka}, {Lainey},
  {LeBouquin}, {Liu}, {Loreggia}, {Makarov}, {Marseille}, {Martayan},
  {Martinez-Rubi}, {Massart}, {Meynadier}, {Mignot}, {Munari}, {Nguyen},
  {Nordlander}, {Ocvirk}, {O'Flaherty}, {Olias Sanz}, {Ortiz}, {Osorio},
  {Oszkiewicz}, {Ouzounis}, {Palmer}, {Park}, {Pasquato}, {Peltzer}, {Peralta},
  {P{\'e}turaud}, {Pieniluoma}, {Pigozzi}, {Poels}, {Prat}, {Prod'homme},
  {Raison}, {Rebordao}, {Risquez}, {Rocca-Volmerange}, {Rosen}, {Ruiz-Fuertes},
  {Russo}, {Sembay}, {Serraller Vizcaino}, {Short}, {Siebert}, {Silva},
  {Sinachopoulos}, {Slezak}, {Soffel}, {Sosnowska}, {Strai{\v{z}}ys}, {ter
  Linden}, {Terrell}, {Theil}, {Tiede}, {Troisi}, {Tsalmantza}, {Tur},
  {Vaccari}, {Vachier}, {Valles}, {Van Hamme}, {Veltz}, {Virtanen}, {Wallut},
  {Wichmann}, {Wilkinson}, {Ziaeepour}, \& {Zschocke}}]{gai16}
{Gaia Collaboration}, {Prusti}, T., {de Bruijne}, J.~H.~J., {et~al.} 2016,
  \aap, 595, A1, \dodoi{10.1051/0004-6361/201629272}

\bibitem[{{Gaia Collaboration} {et~al.}(2021){Gaia Collaboration}, {Brown},
  {Vallenari}, {Prusti}, {de Bruijne}, {Babusiaux}, {Biermann}, {Creevey},
  {Evans}, {Eyer}, {Hutton}, {Jansen}, {Jordi}, {Klioner}, {Lammers},
  {Lindegren}, {Luri}, {Mignard}, {Panem}, {Pourbaix}, {Randich}, {Sartoretti},
  {Soubiran}, {Walton}, {Arenou}, {Bailer-Jones}, {Bastian}, {Cropper},
  {Drimmel}, {Katz}, {Lattanzi}, {van Leeuwen}, {Bakker}, {Cacciari},
  {Casta{\~n}eda}, {De Angeli}, {Ducourant}, {Fabricius}, {Fouesneau},
  {Fr{\'e}mat}, {Guerra}, {Guerrier}, {Guiraud}, {Jean-Antoine Piccolo},
  {Masana}, {Messineo}, {Mowlavi}, {Nicolas}, {Nienartowicz}, {Pailler},
  {Panuzzo}, {Riclet}, {Roux}, {Seabroke}, {Sordo}, {Tanga}, {Th{\'e}venin},
  {Gracia-Abril}, {Portell}, {Teyssier}, {Altmann}, {Andrae}, {Bellas-Velidis},
  {Benson}, {Berthier}, {Blomme}, {Brugaletta}, {Burgess}, {Busso}, {Carry},
  {Cellino}, {Cheek}, {Clementini}, {Damerdji}, {Davidson}, {Delchambre},
  {Dell'Oro}, {Fern{\'a}ndez-Hern{\'a}ndez}, {Galluccio}, {Garc{\'\i}a-Lario},
  {Garcia-Reinaldos}, {Gonz{\'a}lez-N{\'u}{\~n}ez}, {Gosset}, {Haigron},
  {Halbwachs}, {Hambly}, {Harrison}, {Hatzidimitriou}, {Heiter},
  {Hern{\'a}ndez}, {Hestroffer}, {Hodgkin}, {Holl}, {Jan{\ss}en}, {Jevardat de
  Fombelle}, {Jordan}, {Krone-Martins}, {Lanzafame}, {L{\"o}ffler}, {Lorca},
  {Manteiga}, {Marchal}, {Marrese}, {Moitinho}, {Mora}, {Muinonen}, {Osborne},
  {Pancino}, {Pauwels}, {Petit}, {Recio-Blanco}, {Richards}, {Riello},
  {Rimoldini}, {Robin}, {Roegiers}, {Rybizki}, {Sarro}, {Siopis}, {Smith},
  {Sozzetti}, {Ulla}, {Utrilla}, {van Leeuwen}, {van Reeven}, {Abbas}, {Abreu
  Aramburu}, {Accart}, {Aerts}, {Aguado}, {Ajaj}, {Altavilla}, {{\'A}lvarez},
  {{\'A}lvarez Cid-Fuentes}, {Alves}, {Anderson}, {Anglada Varela}, {Antoja},
  {Audard}, {Baines}, {Baker}, {Balaguer-N{\'u}{\~n}ez}, {Balbinot}, {Balog},
  {Barache}, {Barbato}, {Barros}, {Barstow}, {Bartolom{\'e}}, {Bassilana},
  {Bauchet}, {Baudesson-Stella}, {Becciani}, {Bellazzini}, {Bernet}, {Bertone},
  {Bianchi}, {Blanco-Cuaresma}, {Boch}, {Bombrun}, {Bossini}, {Bouquillon},
  {Bragaglia}, {Bramante}, {Breedt}, {Bressan}, {Brouillet}, {Bucciarelli},
  {Burlacu}, {Busonero}, {Butkevich}, {Buzzi}, {Caffau}, {Cancelliere},
  {C{\'a}novas}, {Cantat-Gaudin}, {Carballo}, {Carlucci}, {Carnerero},
  {Carrasco}, {Casamiquela}, {Castellani}, {Castro-Ginard}, {Castro Sampol},
  {Chaoul}, {Charlot}, {Chemin}, {Chiavassa}, {Cioni}, {Comoretto}, {Cooper},
  {Cornez}, {Cowell}, {Crifo}, {Crosta}, {Crowley}, {Dafonte}, {Dapergolas},
  {David}, {David}, {de Laverny}, {De Luise}, {De March}, {De Ridder}, {de
  Souza}, {de Teodoro}, {de Torres}, {del Peloso}, {del Pozo}, {Delbo},
  {Delgado}, {Delgado}, {Delisle}, {Di Matteo}, {Diakite}, {Diener},
  {Distefano}, {Dolding}, {Eappachen}, {Edvardsson}, {Enke}, {Esquej}, {Fabre},
  {Fabrizio}, {Faigler}, {Fedorets}, {Fernique}, {Fienga}, {Figueras},
  {Fouron}, {Fragkoudi}, {Fraile}, {Franke}, {Gai}, {Garabato},
  {Garcia-Gutierrez}, {Garc{\'\i}a-Torres}, {Garofalo}, {Gavras}, {Gerlach},
  {Geyer}, {Giacobbe}, {Gilmore}, {Girona}, {Giuffrida}, {Gomel}, {Gomez},
  {Gonzalez-Santamaria}, {Gonz{\'a}lez-Vidal}, {Granvik},
  {Guti{\'e}rrez-S{\'a}nchez}, {Guy}, {Hauser}, {Haywood}, {Helmi}, {Hidalgo},
  {Hilger}, {H{\l}adczuk}, {Hobbs}, {Holland}, {Huckle}, {Jasniewicz},
  {Jonker}, {Juaristi Campillo}, {Julbe}, {Karbevska}, {Kervella}, {Khanna},
  {Kochoska}, {Kontizas}, {Kordopatis}, {Korn}, {Kostrzewa-Rutkowska},
  {Kruszy{\'n}ska}, {Lambert}, {Lanza}, {Lasne}, {Le Campion}, {Le Fustec},
  {Lebreton}, {Lebzelter}, {Leccia}, {Leclerc}, {Lecoeur-Taibi}, {Liao},
  {Licata}, {Lindstr{\o}m}, {Lister}, {Livanou}, {Lobel}, {Madrero Pardo},
  {Managau}, {Mann}, {Marchant}, {Marconi}, {Marcos Santos}, {Marinoni},
  {Marocco}, {Marshall}, {Martin Polo}, {Mart{\'\i}n-Fleitas}, {Masip},
  {Massari}, {Mastrobuono-Battisti}, {Mazeh}, {McMillan}, {Messina},
  {Michalik}, {Millar}, {Mints}, {Molina}, {Molinaro}, {Moln{\'a}r},
  {Montegriffo}, {Mor}, {Morbidelli}, {Morel}, {Morris}, {Mulone}, {Munoz},
  {Muraveva}, {Murphy}, {Musella}, {Noval}, {Ord{\'e}novic}, {Orr{\`u}},
  {Osinde}, {Pagani}, {Pagano}, {Palaversa}, {Palicio}, {Panahi}, {Pawlak},
  {Pe{\~n}alosa Esteller}, {Penttil{\"a}}, {Piersimoni}, {Pineau}, {Plachy},
  {Plum}, {Poggio}, {Poretti}, {Poujoulet}, {Pr{\v{s}}a}, {Pulone}, {Racero},
  {Ragaini}, {Rainer}, {Raiteri}, {Rambaux}, {Ramos}, {Ramos-Lerate}, {Re
  Fiorentin}, {Regibo}, {Reyl{\'e}}, {Ripepi}, {Riva}, {Rixon}, {Robichon},
  {Robin}, {Roelens}, {Rohrbasser}, {Romero-G{\'o}mez}, {Rowell}, {Royer},
  {Rybicki}, {Sadowski}, {Sagrist{\`a} Sell{\'e}s}, {Sahlmann}, {Salgado},
  {Salguero}, {Samaras}, {Sanchez Gimenez}, {Sanna}, {Santove{\~n}a},
  {Sarasso}, {Schultheis}, {Sciacca}, {Segol}, {Segovia}, {S{\'e}gransan},
  {Semeux}, {Shahaf}, {Siddiqui}, {Siebert}, {Siltala}, {Slezak}, {Smart},
  {Solano}, {Solitro}, {Souami}, {Souchay}, {Spagna}, {Spoto}, {Steele},
  {Steidelm{\"u}ller}, {Stephenson}, {S{\"u}veges}, {Szabados}, {Szegedi-Elek},
  {Taris}, {Tauran}, {Taylor}, {Teixeira}, {Thuillot}, {Tonello}, {Torra},
  {Torra}, {Turon}, {Unger}, {Vaillant}, {van Dillen}, {Vanel}, {Vecchiato},
  {Viala}, {Vicente}, {Voutsinas}, {Weiler}, {Wevers}, {Wyrzykowski}, {Yoldas},
  {Yvard}, {Zhao}, {Zorec}, {Zucker}, {Zurbach}, \& {Zwitter}}]{gai21}
{Gaia Collaboration}, {Brown}, A.~G.~A., {Vallenari}, A., {et~al.} 2021, \aap,
  649, A1, \dodoi{10.1051/0004-6361/202039657}

\bibitem[{{Garc{\'\i}a} {et~al.}(2018){Garc{\'\i}a}, {Fogantini}, {Chaty}, \&
  {Combi}}]{gar18}
{Garc{\'\i}a}, F., {Fogantini}, F.~A., {Chaty}, S., \& {Combi}, J.~A. 2018,
  \aap, 618, A61, \dodoi{10.1051/0004-6361/201833365}

\bibitem[{{Gim{\'e}nez-Garc{\'\i}a} {et~al.}(2015){Gim{\'e}nez-Garc{\'\i}a},
  {Torrej{\'o}n}, {Eikmann}, {Mart{\'\i}nez-N{\'u}{\~n}ez}, {Oskinova},
  {Rodes-Roca}, \& {Bernab{\'e}u}}]{gim15}
{Gim{\'e}nez-Garc{\'\i}a}, A., {Torrej{\'o}n}, J.~M., {Eikmann}, W., {et~al.}
  2015, \aap, 576, A108, \dodoi{10.1051/0004-6361/201425004}

\bibitem[{{Grebenev} \& {Sunyaev}(2007)}]{gre07}
{Grebenev}, S.~A., \& {Sunyaev}, R.~A. 2007, Astronomy Letters, 33, 149,
  \dodoi{10.1134/S1063773707030024}

\bibitem[{{Hainich} {et~al.}(2020){Hainich}, {Oskinova}, {Torrej{\'o}n},
  {Fuerst}, {Bodaghee}, {Shenar}, {Sander}, {Todt}, {Spetzer}, \&
  {Hamann}}]{hai20}
{Hainich}, R., {Oskinova}, L.~M., {Torrej{\'o}n}, J.~M., {et~al.} 2020, \aap,
  634, A49, \dodoi{10.1051/0004-6361/201935498}

\bibitem[{{Horne} \& {Baliunas}(1986)}]{hor86}
{Horne}, J.~H., \& {Baliunas}, S.~L. 1986, \apj, 302, 757,
  \dodoi{10.1086/164037}

\bibitem[{{Illarionov} \& {Sunyaev}(1975)}]{ill75}
{Illarionov}, A.~F., \& {Sunyaev}, R.~A. 1975, \aap, 39, 185

\bibitem[{{in't Zand}(2005)}]{int05}
{in't Zand}, J.~J.~M. 2005, \aap, 441, L1, \dodoi{10.1051/0004-6361:200500162}

\bibitem[{{Jenke} {et~al.}(2012){Jenke}, {Finger}, {Wilson-Hodge}, \&
  {Camero-Arranz}}]{jen12}
{Jenke}, P.~A., {Finger}, M.~H., {Wilson-Hodge}, C.~A., \& {Camero-Arranz}, A.
  2012, \apj, 759, 124, \dodoi{10.1088/0004-637X/759/2/124}

\bibitem[{{Kretschmar} {et~al.}(2019){Kretschmar}, {F{\"u}rst}, {Sidoli},
  {Bozzo}, {Alfonso-Garz{\'o}n}, {Bodaghee}, {Chaty}, {Chernyakova},
  {Ferrigno}, {Manousakis}, {Negueruela}, {Postnov}, {Paizis}, {Reig},
  {Rodes-Roca}, {Tsygankov}, {Bird}, {Bissinger n{\'e} K{\"u}hnel}, {Blay},
  {Caballero}, {Coe}, {Domingo}, {Doroshenko}, {Ducci}, {Falanga}, {Grebenev},
  {Grinberg}, {Hemphill}, {Kreykenbohm}, {Kreykenbohm n{\'e} Fritz}, {Li},
  {Lutovinov}, {Mart{\'\i}nez-N{\'u}{\~n}ez}, {Mas-Hesse}, {Masetti},
  {McBride}, {Neronov}, {Pottschmidt}, {Rodriguez}, {Romano}, {Rothschild},
  {Santangelo}, {Sguera}, {Staubert}, {Tomsick}, {Torrej{\'o}n}, {Torres},
  {Walter}, {Wilms}, {Wilson-Hodge}, \& {Zhang}}]{kre19}
{Kretschmar}, P., {F{\"u}rst}, F., {Sidoli}, L., {et~al.} 2019, \nar, 86,
  101546, \dodoi{10.1016/j.newar.2020.101546}

\bibitem[{{Krimm} {et~al.}(2013){Krimm}, {Holland}, {Corbet}, {Pearlman},
  {Romano}, {Kennea}, {Bloom}, {Barthelmy}, {Baumgartner}, {Cummings},
  {Gehrels}, {Lien}, {Markwardt}, {Palmer}, {Sakamoto}, {Stamatikos}, \&
  {Ukwatta}}]{kri13}
{Krimm}, H.~A., {Holland}, S.~T., {Corbet}, R.~H.~D., {et~al.} 2013, \apjs,
  209, 14, \dodoi{10.1088/0067-0049/209/1/14}

\bibitem[{{Krivonos} {et~al.}(2022){Krivonos}, {Sazonov}, {Kuznetsova},
  {Lutovinov}, {Mereminskiy}, \& {Tsygankov}}]{kri22}
{Krivonos}, R.~A., {Sazonov}, S.~Y., {Kuznetsova}, E.~A., {et~al.} 2022,
  \mnras, 510, 4796, \dodoi{10.1093/mnras/stab3751}

\bibitem[{{Leahy}(1987)}]{lea87}
{Leahy}, D.~A. 1987, \aap, 180, 275

\bibitem[{{Levine} {et~al.}(2011){Levine}, {Bradt}, {Chakrabarty}, {Corbet}, \&
  {Harris}}]{lev11}
{Levine}, A.~M., {Bradt}, H.~V., {Chakrabarty}, D., {Corbet}, R. H.~D., \&
  {Harris}, R.~J. 2011, \apjs, 196, 6, \dodoi{10.1088/0067-0049/196/1/6}

\bibitem[{{Lomb}(1976)}]{lom76}
{Lomb}, N.~R. 1976, \apss, 39, 447, \dodoi{10.1007/BF00648343}

\bibitem[{{Lutovinov} {et~al.}(2005){Lutovinov}, {Rodriguez}, {Revnivtsev}, \&
  {Shtykovskiy}}]{lut05}
{Lutovinov}, A., {Rodriguez}, J., {Revnivtsev}, M., \& {Shtykovskiy}, P. 2005,
  \aap, 433, L41, \dodoi{10.1051/0004-6361:200500092}

\bibitem[{{Lutovinov} {et~al.}(2017){Lutovinov}, {Tsygankov}, {Postnov},
  {Krivonos}, {Molkov}, \& {Tomsick}}]{lut17}
{Lutovinov}, A.~A., {Tsygankov}, S.~S., {Postnov}, K.~A., {et~al.} 2017,
  \mnras, 466, 593, \dodoi{10.1093/mnras/stw3058}

\bibitem[{{Magdziarz} \& {Zdziarski}(1995)}]{mag95}
{Magdziarz}, P., \& {Zdziarski}, A.~A. 1995, \mnras, 273, 837,
  \dodoi{10.1093/mnras/273.3.837}

\bibitem[{{Makishima} {et~al.}(1999){Makishima}, {Mihara}, {Nagase}, \&
  {Tanaka}}]{mak99}
{Makishima}, K., {Mihara}, T., {Nagase}, F., \& {Tanaka}, Y. 1999, \apj, 525,
  978, \dodoi{10.1086/307912}

\bibitem[{{Makishima} {et~al.}(1990){Makishima}, {Mihara}, {Ishida}, {Ohashi},
  {Sakao}, {Tashiro}, {Tsuru}, {Kii}, {Makino}, {Murakami}, {Nagase}, {Tanaka},
  {Kunieda}, {Tawara}, {Kitamoto}, {Miyamoto}, {Yoshida}, \& {Turner}}]{mak90}
{Makishima}, K., {Mihara}, T., {Ishida}, M., {et~al.} 1990, \apjl, 365, L59,
  \dodoi{10.1086/185888}

\bibitem[{{Manousakis} {et~al.}(2012){Manousakis}, {Walter}, \&
  {Blondin}}]{man12}
{Manousakis}, A., {Walter}, R., \& {Blondin}, J.~M. 2012, \aap, 547, A20,
  \dodoi{10.1051/0004-6361/201219717}

\bibitem[{MATLAB(2021)}]{mat21}
MATLAB. 2021, version 9.11.0 (R2021b) (Natick, Massachusetts: The MathWorks
  Inc.)

\bibitem[{{Matt} \& {Guainazzi}(2003)}]{mat03}
{Matt}, G., \& {Guainazzi}, M. 2003, \mnras, 341, L13,
  \dodoi{10.1046/j.1365-8711.2003.06658.x}

\bibitem[{{Mihara} {et~al.}(1990){Mihara}, {Makishima}, {Ohashi}, {Sakao}, \&
  {Tashiro}}]{mih90}
{Mihara}, T., {Makishima}, K., {Ohashi}, T., {Sakao}, T., \& {Tashiro}, M.
  1990, \nat, 346, 250, \dodoi{10.1038/346250a0}

\bibitem[{{Mushtukov} \& {Tsygankov}(2022)}]{mus22}
{Mushtukov}, A., \& {Tsygankov}, S. 2022, arXiv e-prints, arXiv:2204.14185.
\newblock \doarXiv{2204.14185}

\bibitem[{{Nagase}(1989)}]{nag89}
{Nagase}, F. 1989, \pasj, 41, 1

\bibitem[{{Nasa High Energy Astrophysics Science Archive Research Center
  (Heasarc)}(2014)}]{hea14}
{Nasa High Energy Astrophysics Science Archive Research Center (Heasarc)}.
  2014, {HEAsoft: Unified Release of FTOOLS and XANADU}, Astrophysics Source
  Code Library, record ascl:1408.004.
\newblock \doeprint{1408.004}

\bibitem[{{Negueruela} {et~al.}(2006){Negueruela}, {Smith}, {Reig}, {Chaty}, \&
  {Torrej{\'o}n}}]{neg06}
{Negueruela}, I., {Smith}, D.~M., {Reig}, P., {Chaty}, S., \& {Torrej{\'o}n},
  J.~M. 2006, in ESA Special Publication, Vol. 604, The X-ray Universe 2005,
  ed. A.~{Wilson}, 165.
\newblock \doarXiv{astro-ph/0511088}

\bibitem[{{Orlandini} {et~al.}(2012){Orlandini}, {Frontera}, {Masetti},
  {Sguera}, \& {Sidoli}}]{orl12}
{Orlandini}, M., {Frontera}, F., {Masetti}, N., {Sguera}, V., \& {Sidoli}, L.
  2012, \apj, 748, 86, \dodoi{10.1088/0004-637X/748/2/86}

\bibitem[{{Oskinova} {et~al.}(2012){Oskinova}, {Feldmeier}, \&
  {Kretschmar}}]{osk12}
{Oskinova}, L.~M., {Feldmeier}, A., \& {Kretschmar}, P. 2012, \mnras, 421,
  2820, \dodoi{10.1111/j.1365-2966.2012.20507.x}

\bibitem[{{Patel} {et~al.}(2004){Patel}, {Kouveliotou}, {Tennant}, {Woods},
  {King}, {Finger}, {Ubertini}, {Winkler}, {Courvoisier}, {van der Klis},
  {Wachter}, {Gaensler}, \& {Phillips}}]{pat04}
{Patel}, S.~K., {Kouveliotou}, C., {Tennant}, A., {et~al.} 2004, \apjl, 602,
  L45, \dodoi{10.1086/382210}

\bibitem[{{Patel} {et~al.}(2007){Patel}, {Zurita}, {Del Santo}, {Finger},
  {Kouveliotou}, {Eichler}, {G{\"o}{\v{g}}{\"u}{\c{s}}}, {Ubertini}, {Walter},
  {Woods}, {Wilson}, {Wachter}, \& {Bazzano}}]{pat07}
{Patel}, S.~K., {Zurita}, J., {Del Santo}, M., {et~al.} 2007, \apj, 657, 994,
  \dodoi{10.1086/510374}

\bibitem[{{Pradhan} {et~al.}(2018){Pradhan}, {Bozzo}, \& {Paul}}]{pra18}
{Pradhan}, P., {Bozzo}, E., \& {Paul}, B. 2018, \aap, 610, A50,
  \dodoi{10.1051/0004-6361/201731487}

\bibitem[{{Press} \& {Rybicki}(1989)}]{pre89}
{Press}, W.~H., \& {Rybicki}, G.~B. 1989, \apj, 338, 277,
  \dodoi{10.1086/167197}

\bibitem[{{Protassov} {et~al.}(2002){Protassov}, {van Dyk}, {Connors},
  {Kashyap}, \& {Siemiginowska}}]{pro02}
{Protassov}, R., {van Dyk}, D.~A., {Connors}, A., {Kashyap}, V.~L., \&
  {Siemiginowska}, A. 2002, \apj, 571, 545, \dodoi{10.1086/339856}

\bibitem[{{Rahoui} {et~al.}(2008){Rahoui}, {Chaty}, {Lagage}, \&
  {Pantin}}]{rah08}
{Rahoui}, F., {Chaty}, S., {Lagage}, P.~O., \& {Pantin}, E. 2008, \aap, 484,
  801, \dodoi{10.1051/0004-6361:20078774}

\bibitem[{{Rodriguez} {et~al.}(2003){Rodriguez}, {Tomsick}, {Foschini},
  {Walter}, {Goldwurm}, {Corbel}, \& {Kaaret}}]{rod03}
{Rodriguez}, J., {Tomsick}, J.~A., {Foschini}, L., {et~al.} 2003, \aap, 407,
  L41, \dodoi{10.1051/0004-6361:20031093}

\bibitem[{{Rodriguez} {et~al.}(2006){Rodriguez}, {Bodaghee}, {Kaaret},
  {Tomsick}, {Kuulkers}, {Malaguti}, {Petrucci}, {Cabanac}, {Chernyakova},
  {Corbel}, {Deluit}, {Di Cocco}, {Ebisawa}, {Goldwurm}, {Henri}, {Lebrun},
  {Paizis}, {Walter}, \& {Foschini}}]{rod06}
{Rodriguez}, J., {Bodaghee}, A., {Kaaret}, P., {et~al.} 2006, \mnras, 366, 274,
  \dodoi{10.1111/j.1365-2966.2005.09855.x}

\bibitem[{{Romano} {et~al.}(2014){Romano}, {Krimm}, {Palmer}, {Ducci},
  {Esposito}, {Vercellone}, {Evans}, {Guidorzi}, {Mangano}, {Kennea},
  {Barthelmy}, {Burrows}, \& {Gehrels}}]{rom14}
{Romano}, P., {Krimm}, H.~A., {Palmer}, D.~M., {et~al.} 2014, \aap, 562, A2,
  \dodoi{10.1051/0004-6361/201322516}

\bibitem[{{Romano} {et~al.}(2015){Romano}, {Bozzo}, {Mangano}, {Esposito},
  {Israel}, {Tiengo}, {Campana}, {Ducci}, {Ferrigno}, \& {Kennea}}]{rom15}
{Romano}, P., {Bozzo}, E., {Mangano}, V., {et~al.} 2015, \aap, 576, L4,
  \dodoi{10.1051/0004-6361/201525749}

\bibitem[{{Sartore} {et~al.}(2015){Sartore}, {Jourdain}, \& {Roques}}]{sar15}
{Sartore}, N., {Jourdain}, E., \& {Roques}, J.~P. 2015, \apj, 806, 193,
  \dodoi{10.1088/0004-637X/806/2/193}

\bibitem[{{Scargle}(1982)}]{sca82}
{Scargle}, J.~D. 1982, \apj, 263, 835, \dodoi{10.1086/160554}

\bibitem[{{Sch{\"o}nherr} {et~al.}(2014){Sch{\"o}nherr}, {Schwarm}, {Falkner},
  {Dauser}, {Ferrigno}, {K{\"u}hnel}, {Klochkov}, {Kretschmar}, {Becker},
  {Wolff}, {Pottschmidt}, {Falanga}, {Kreykenbohm}, {F{\"u}rst}, {Staubert}, \&
  {Wilms}}]{sch14}
{Sch{\"o}nherr}, G., {Schwarm}, F.~W., {Falkner}, S., {et~al.} 2014, \aap, 564,
  L8, \dodoi{10.1051/0004-6361/201322448}

\bibitem[{{Schwarm} {et~al.}(2017{\natexlab{a}}){Schwarm}, {Sch{\"o}nherr},
  {Falkner}, {Pottschmidt}, {Wolff}, {Becker}, {Sokolova-Lapa}, {Klochkov},
  {Ferrigno}, {F{\"u}rst}, {Hemphill}, {Marcu-Cheatham}, {Dauser}, \&
  {Wilms}}]{sch17a}
{Schwarm}, F.~W., {Sch{\"o}nherr}, G., {Falkner}, S., {et~al.}
  2017{\natexlab{a}}, \aap, 597, A3, \dodoi{10.1051/0004-6361/201629352}

\bibitem[{{Schwarm} {et~al.}(2017{\natexlab{b}}){Schwarm}, {Ballhausen},
  {Falkner}, {Sch{\"o}nherr}, {Pottschmidt}, {Wolff}, {Becker}, {F{\"u}rst},
  {Marcu-Cheatham}, {Hemphill}, {Sokolova-Lapa}, {Dauser}, {Klochkov},
  {Ferrigno}, \& {Wilms}}]{sch17b}
{Schwarm}, F.~W., {Ballhausen}, R., {Falkner}, S., {et~al.} 2017{\natexlab{b}},
  \aap, 601, A99, \dodoi{10.1051/0004-6361/201630250}

\bibitem[{{Shakura} {et~al.}(2012){Shakura}, {Postnov}, {Kochetkova}, \&
  {Hjalmarsdotter}}]{sha12}
{Shakura}, N., {Postnov}, K., {Kochetkova}, A., \& {Hjalmarsdotter}, L. 2012,
  \mnras, 420, 216, \dodoi{10.1111/j.1365-2966.2011.20026.x}

\bibitem[{{Sidoli} \& {Paizis}(2018)}]{sid18}
{Sidoli}, L., \& {Paizis}, A. 2018, \mnras, 481, 2779,
  \dodoi{10.1093/mnras/sty2428}

\bibitem[{{Sidoli} {et~al.}(2016){Sidoli}, {Paizis}, \& {Postnov}}]{sid16}
{Sidoli}, L., {Paizis}, A., \& {Postnov}, K. 2016, \mnras, 457, 3693,
  \dodoi{10.1093/mnras/stw237}

\bibitem[{{Soong} {et~al.}(1990){Soong}, {Gruber}, {Peterson}, \&
  {Rothschild}}]{soo90}
{Soong}, Y., {Gruber}, D.~E., {Peterson}, L.~E., \& {Rothschild}, R.~E. 1990,
  \apj, 348, 641, \dodoi{10.1086/168272}

\bibitem[{{Staubert} {et~al.}(2019){Staubert}, {Tr{\"u}mper}, {Kendziorra},
  {Klochkov}, {Postnov}, {Kretschmar}, {Pottschmidt}, {Haberl}, {Rothschild},
  {Santangelo}, {Wilms}, {Kreykenbohm}, \& {F{\"u}rst}}]{sta19}
{Staubert}, R., {Tr{\"u}mper}, J., {Kendziorra}, E., {et~al.} 2019, \aap, 622,
  A61, \dodoi{10.1051/0004-6361/201834479}

\bibitem[{{Suchy} {et~al.}(2012){Suchy}, {F{\"u}rst}, {Pottschmidt},
  {Caballero}, {Kreykenbohm}, {Wilms}, {Markowitz}, \& {Rothschild}}]{suc12}
{Suchy}, S., {F{\"u}rst}, F., {Pottschmidt}, K., {et~al.} 2012, \apj, 745, 124,
  \dodoi{10.1088/0004-637X/745/2/124}

\bibitem[{{Sugizaki} {et~al.}(2001){Sugizaki}, {Mitsuda}, {Kaneda},
  {Matsuzaki}, {Yamauchi}, \& {Koyama}}]{sug01}
{Sugizaki}, M., {Mitsuda}, K., {Kaneda}, H., {et~al.} 2001, \apjs, 134, 77,
  \dodoi{10.1086/320358}

\bibitem[{{Tanaka}(1986)}]{tan86}
{Tanaka}, Y. 1986, in IAU Colloq. 89: Radiation Hydrodynamics in Stars and
  Compact Objects, ed. D.~{Mihalas} \& K.-H.~A. {Winkler}, Vol. 255, 198,
  \dodoi{10.1007/3-540-16764-1\_12}

\bibitem[{{Tendulkar} {et~al.}(2014){Tendulkar}, {F{\"u}rst}, {Pottschmidt},
  {Bachetti}, {Bhalerao}, {Boggs}, {Christensen}, {Craig}, {Hailey},
  {Harrison}, {Stern}, {Tomsick}, {Walton}, \& {Zhang}}]{shr14}
{Tendulkar}, S.~P., {F{\"u}rst}, F., {Pottschmidt}, K., {et~al.} 2014, \apj,
  795, 154, \dodoi{10.1088/0004-637X/795/2/154}

\bibitem[{{Titarchuk}(1994)}]{tit94}
{Titarchuk}, L. 1994, \apj, 434, 570, \dodoi{10.1086/174760}

\bibitem[{{Tomsick} {et~al.}(2003){Tomsick}, {Lingenfelter}, {Walter},
  {Rodriguez}, {Goldwurm}, {Corbel}, \& {Kaaret}}]{tom03}
{Tomsick}, J.~A., {Lingenfelter}, R., {Walter}, R., {et~al.} 2003, \iaucirc,
  8076, 1

\bibitem[{{Truemper} {et~al.}(1978){Truemper}, {Pietsch}, {Reppin}, {Voges},
  {Staubert}, \& {Kendziorra}}]{tru78}
{Truemper}, J., {Pietsch}, W., {Reppin}, C., {et~al.} 1978, \apjl, 219, L105,
  \dodoi{10.1086/182617}

\bibitem[{{Vasco} {et~al.}(2013){Vasco}, {Staubert}, {Klochkov}, {Santangelo},
  {Shakura}, \& {Postnov}}]{vas13}
{Vasco}, D., {Staubert}, R., {Klochkov}, D., {et~al.} 2013, \aap, 550, A111,
  \dodoi{10.1051/0004-6361/201220181}

\bibitem[{{Verner} {et~al.}(1996){Verner}, {Ferland}, {Korista}, \&
  {Yakovlev}}]{ver96}
{Verner}, D.~A., {Ferland}, G.~J., {Korista}, K.~T., \& {Yakovlev}, D.~G. 1996,
  \apj, 465, 487, \dodoi{10.1086/177435}

\bibitem[{{Walter} {et~al.}(2015){Walter}, {Lutovinov}, {Bozzo}, \&
  {Tsygankov}}]{wal15}
{Walter}, R., {Lutovinov}, A.~A., {Bozzo}, E., \& {Tsygankov}, S.~S. 2015,
  \aapr, 23, 2, \dodoi{10.1007/s00159-015-0082-6}

\bibitem[{{Walter} {et~al.}(2003){Walter}, {Rodriguez}, {Foschini}, {de Plaa},
  {Corbel}, {Courvoisier}, {den Hartog}, {Lebrun}, {Parmar}, {Tomsick}, \&
  {Ubertini}}]{wal03}
{Walter}, R., {Rodriguez}, J., {Foschini}, L., {et~al.} 2003, \aap, 411, L427,
  \dodoi{10.1051/0004-6361:20031369}

\bibitem[{{Webb} {et~al.}(2020){Webb}, {Coriat}, {Traulsen}, {Ballet}, {Motch},
  {Carrera}, {Koliopanos}, {Authier}, {de la Calle}, {Ceballos}, {Colomo},
  {Chuard}, {Freyberg}, {Garcia}, {Kolehmainen}, {Lamer}, {Lin}, {Maggi},
  {Michel}, {Page}, {Page}, {Perea-Calderon}, {Pineau}, {Rodriguez}, {Rosen},
  {Santos Lleo}, {Saxton}, {Schwope}, {Tom{\'a}s}, {Watson}, \&
  {Zakardjian}}]{web20}
{Webb}, N.~A., {Coriat}, M., {Traulsen}, I., {et~al.} 2020, \aap, 641, A136,
  \dodoi{10.1051/0004-6361/201937353}

\bibitem[{{White} {et~al.}(1983){White}, {Swank}, \& {Holt}}]{whi83}
{White}, N.~E., {Swank}, J.~H., \& {Holt}, S.~S. 1983, \apj, 270, 711,
  \dodoi{10.1086/161162}

\bibitem[{{Wilms} {et~al.}(2000){Wilms}, {Allen}, \& {McCray}}]{wil00}
{Wilms}, J., {Allen}, A., \& {McCray}, R. 2000, \apj, 542, 914,
  \dodoi{10.1086/317016}

\bibitem[{{Zdziarski} {et~al.}(1996){Zdziarski}, {Johnson}, \&
  {Magdziarz}}]{zdz96}
{Zdziarski}, A.~A., {Johnson}, W.~N., \& {Magdziarz}, P. 1996, \mnras, 283,
  193, \dodoi{10.1093/mnras/283.1.193}

\bibitem[{{Zurita Heras} {et~al.}(2006){Zurita Heras}, {De Cesare}, {Walter},
  {Bodaghee}, {B{\'e}langer}, {Courvoisier}, {Shaw}, \& {Stephen}}]{zur06}
{Zurita Heras}, J.~A., {De Cesare}, G., {Walter}, R., {et~al.} 2006, \aap, 448,
  261, \dodoi{10.1051/0004-6361:20053876}

\bibitem[{{{\.Z}ycki} {et~al.}(1999){{\.Z}ycki}, {Done}, \& {Smith}}]{zyc99}
{{\.Z}ycki}, P.~T., {Done}, C., \& {Smith}, D.~A. 1999, \mnras, 309, 561,
  \dodoi{10.1046/j.1365-8711.1999.02885.x}

\end{thebibliography}
\bibliographystyle{aasjournal}



\end{document}